%% file: main.tex
\documentclass{article}

\newif\ifarxiv
\arxivtrue

\ifarxiv
  \usepackage[preprint]{colm2026_conference}
\else
  \usepackage[submission]{colm2026_conference}
\fi

\usepackage[T1]{fontenc}
\usepackage{microtype}
\usepackage{hyperref}
\usepackage{url}
\usepackage{booktabs}
\usepackage{lineno}

\definecolor{darkblue}{rgb}{0, 0, 0.5}
\hypersetup{colorlinks=true, citecolor=darkblue, linkcolor=darkblue, urlcolor=darkblue}

\usepackage{graphicx}
\usepackage{subcaption}
\usepackage{multirow}
\usepackage{makecell}
\usepackage{siunitx}
\usepackage{tabularx}
\usepackage{float}

\usepackage{enumitem}
\usepackage[dvipsnames,table]{xcolor}
\definecolor{lightgray}{gray}{0.93}

\usepackage{amsmath}
\usepackage{csquotes}
\usepackage{xspace}
\usepackage{xfp}

\usepackage{algorithm}
\usepackage{algpseudocode}

\usepackage{listings}
\usepackage{comment}
\newtoggle{comment}
\toggletrue{comment}

\newcommand{\todo}[2][]{\textcolor{red}{\textbf{TODO}%
\ifx&#1&\else\ [#1]\fi: #2}}

\newcommand{\ci}[3]{$#1_{\pm \fpeval{max(#2, #3)}}$}

\usepackage[capitalize,noabbrev]{cleveref}
\crefname{appendix}{Appendix}{Appendices}

\newtoggle{anonymous}
\ifarxiv
  \togglefalse{anonymous}
\else
  \toggletrue{anonymous}
\fi

\iftoggle{anonymous}{
  \newcommand{\sys}{Nora\xspace}
  \newcommand{\dataset}{AI Research Assistant Dataset (ARAD)\xspace}
  \newcommand{\pf}{PS\xspace}
  \newcommand{\sqa}{SciQA\xspace}
  \newcommand{\sTwo}{S2\xspace}
  \newcommand{\pfFull}{Paper Search\xspace}
  \newcommand{\sqaFull}{Science Question Answering\xspace}
}{
  \newcommand{\sys}{Asta\xspace}
  \newcommand{\dataset}{Asta Interaction Dataset (AID)\xspace}
  \newcommand{\pf}{PF\xspace}
  \newcommand{\sqa}{SQA\xspace}
  \newcommand{\sTwo}{S2\xspace}
  \newcommand{\pfFull}{PaperFinder\xspace}
  \newcommand{\sqaFull}{ScholarQA\xspace}
}

\newcommand{\singleQueryDesc}{single-query}
\newcommand{\inexperiencedDesc}{inexperienced}
\newcommand{\experiencedDesc}{experienced}

\title{Understanding Usage and Engagement in AI-Powered Scientific Research Tools: The Asta Interaction Dataset}

\ifarxiv
  \author{\mdseries Dany Haddad\textsuperscript{*\dag},
    Dan Bareket\textsuperscript{*\dag},
    Joseph Chee Chang\textsuperscript{*\dag},
    Jay DeYoung\textsuperscript{*\dag}, \\
    Jena D.\ Hwang\textsuperscript{*\dag},
    Uri Katz\textsuperscript{*\dag},
    Mark Polak\textsuperscript{*\dag},
    Sangho Suh\textsuperscript{*\dag},
    Harshit Surana\textsuperscript{*\dag},
    Aryeh Tiktinsky\textsuperscript{*\dag}, \\\\
    Shriya Atmakuri\textsuperscript{*},
    Jonathan Bragg\textsuperscript{*},
    Mike D'Arcy\textsuperscript{*},
    Sergey Feldman\textsuperscript{*},
    Amal Hassan-Ali\textsuperscript{*}, \\
    Rubén Lozano\textsuperscript{*},
    Bodhisattwa Prasad Majumder\textsuperscript{*},
    Charles McGrady\textsuperscript{*},
    Amanpreet Singh\textsuperscript{*},
    Brooke Vlahos\textsuperscript{*}, \\\\
    Yoav Goldberg\textsuperscript{*\dag},
    Doug Downey\textsuperscript{*\dag} \\
    \textsuperscript{*}Allen Institute for AI \quad \textsuperscript{\dag}Core contributors
  }
\fi

\begin{document}

\ifcolmsubmission
\linenumbers
\fi

\maketitle

\begin{abstract}
AI-powered scientific research tools are rapidly being integrated
into research workflows, yet the field lacks a clear lens into how
researchers use these systems in real-world settings.
We present and analyze the \textit{Asta Interaction Dataset}, a large-scale
resource comprising over 200{,}000 user queries and interaction logs
from two deployed tools (a literature discovery interface and a
scientific question-answering interface) within an LLM-powered
retrieval-augmented generation platform.
Using this dataset, we characterize query patterns, engagement
behaviors, and how usage evolves with experience.
We find that users submit longer and more complex queries than
in traditional search, and treat the system as
a collaborative research partner, delegating tasks such as drafting
content and identifying research gaps.
Users treat generated responses as persistent artifacts, revisiting
and navigating among outputs and cited evidence in non-linear ways.
With experience, users issue more targeted queries and engage more
deeply with supporting citations, although keyword-style queries
persist even among experienced users.
We release the anonymized dataset and analysis with a new query intent taxonomy to
inform future designs of real-world AI research assistants and to support
realistic evaluation.
\end{abstract}

\input{intro}
\input{system}
\input{methods}
\input{results}
\input{related}
\input{discussion}
\bibliography{references}
\bibliographystyle{colm2026_conference}

\input{appendix_methods}

\input{appendix_analysis}
\input{appendix_results}
\input{appendix_dataset}
\input{appendix_taxonomies}
\input{appendix_prompts}

\end{document}

%% file: intro.tex
\section{Introduction}
\label{sec:intro}

AI-powered assistants for scientific question answering and literature
discovery are increasingly deployed in both academic and commercial
settings~\citep{xu2025deepresearchsurvey}, rapidly becoming part of
day-to-day scientific workflows for tasks ranging from paper discovery
to literature reviews and experimental planning~\citep{liao2024llms}. Commercial systems
typically combine retrieval over large scholarly corpora with LLM-based synthesis,
and include general-purpose AI search engines 
\citep[e.g,][]{perplexity2024,youcom2024}; deep research agents
\citep[e.g,][]{google2024deepresearch, openai2025deepresearch,
  anthropic2025research}; and science-focused platforms
\citep[e.g,][]{elicit2024,skarlinski2024paperqa2}.

Despite rapid adoption, we still lack a clear picture of what researchers actually do with these systems: Do they use them like search engines? As writing assistants? As collaborators? Or as something else entirely? Existing studies typically report aggregate statistics derived from proprietary logs~\citep{chatterji2025how, phang2025investigating, handa2025economictasksperformedai}. To our knowledge, no publicly available large-scale dataset of real-world user interactions with deployed AI-powered scientific research tools exists.

We address this gap by releasing and analyzing the \textit{Asta Interaction Dataset}, a large-scale anonymized
user behavior log from \sys{} \citep{singh2025ai2scholarqaorganized,paperfinder2025}.
\sys{} exemplifies the emerging class of AI research assistants,
and integrates with the academic search engine Semantic Scholar to provide two AI-powered interfaces: \pf{}
(\pfFull, a paper search interface) and \sqa{}
(\sqaFull, a scientific question-answering interface).
Our dataset includes over 200,000 queries and associated clickstream logs, forming
the first public dataset of real-world user interactions
with a deployed AI-powered scientific research tool.

We address two interrelated \textbf{research
  questions}: \textbf{RQ1:} How do researchers formulate {\em information needs} when interacting with LLM-based retrieval and synthesis systems, how does it differ from traditional search, and how do these behaviors evolve with experience? \textbf{RQ2:} How do
users consume and navigate AI-generated research {\em reports}, and what does this reveal about how design choices shape engagement?
Across both interfaces, we observe a shift from search-like behavior toward collaborative use: users issue longer, task-oriented queries and delegate higher-level research activities such as drafting and gap identification.  We structure these analyses
within a new taxonomy of query intent, phrasing, and criteria.

\textbf{Contributions:} (1) We publicly release a large-scale dataset
of over 200,000 real-world user queries and interaction logs from deployed
AI-powered research tools. (2) We provide an analysis of query
patterns and user engagement and
how usage behaviors evolve over time. (3) We introduce a multidimensional query taxonomy (intent, phrasing, criteria) tailored to AI research assistants.


%% file: system.tex
\section{System description}
\label{sec:system}

\textbf{Interfaces.} \sys exposes two tools:
\pf{} (a paper-finding interface that returns a ranked list of papers
with lightweight synthesis) and \sqa{} (a scientific
question-answering interface that produces a structured report with
citations). As a comparative baseline, we also analyze queries
performed on Semantic Scholar\footnote{\url{https://semanticscholar.org}} (\sTwo{}), a traditional academic search website.

\begin{figure*}[t]
  \centering
  \begin{subfigure}[t]{0.48\linewidth}
    \centering
    \includegraphics[width=\linewidth]{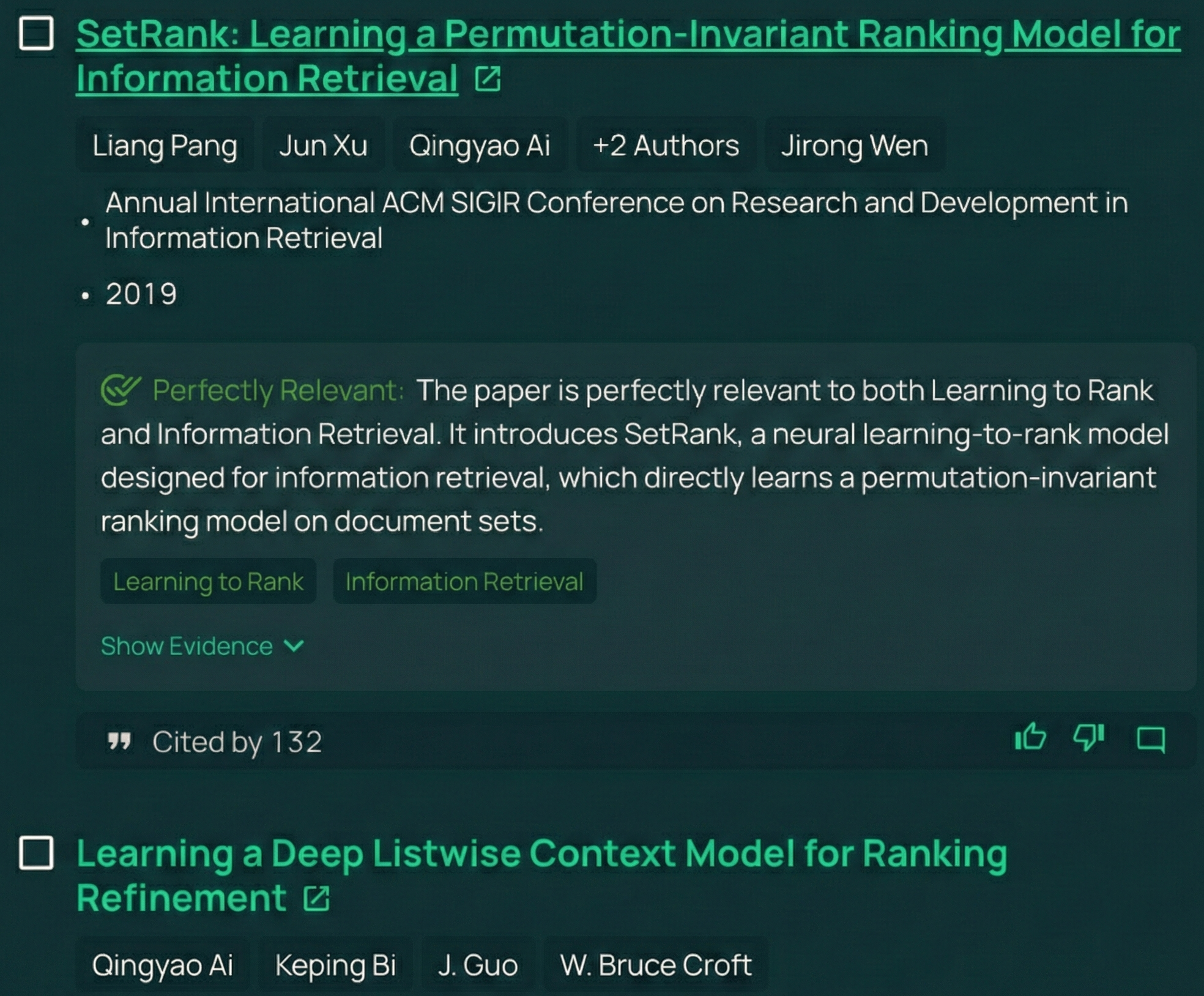}
    \caption{\pf{} interface shows a ranked list of papers with per-item actions
      and filters.}
    \label{fig:pf-mockup}
  \end{subfigure}
  \hfill
  \begin{subfigure}[t]{0.48\linewidth}
    \centering
    \includegraphics[width=\linewidth]{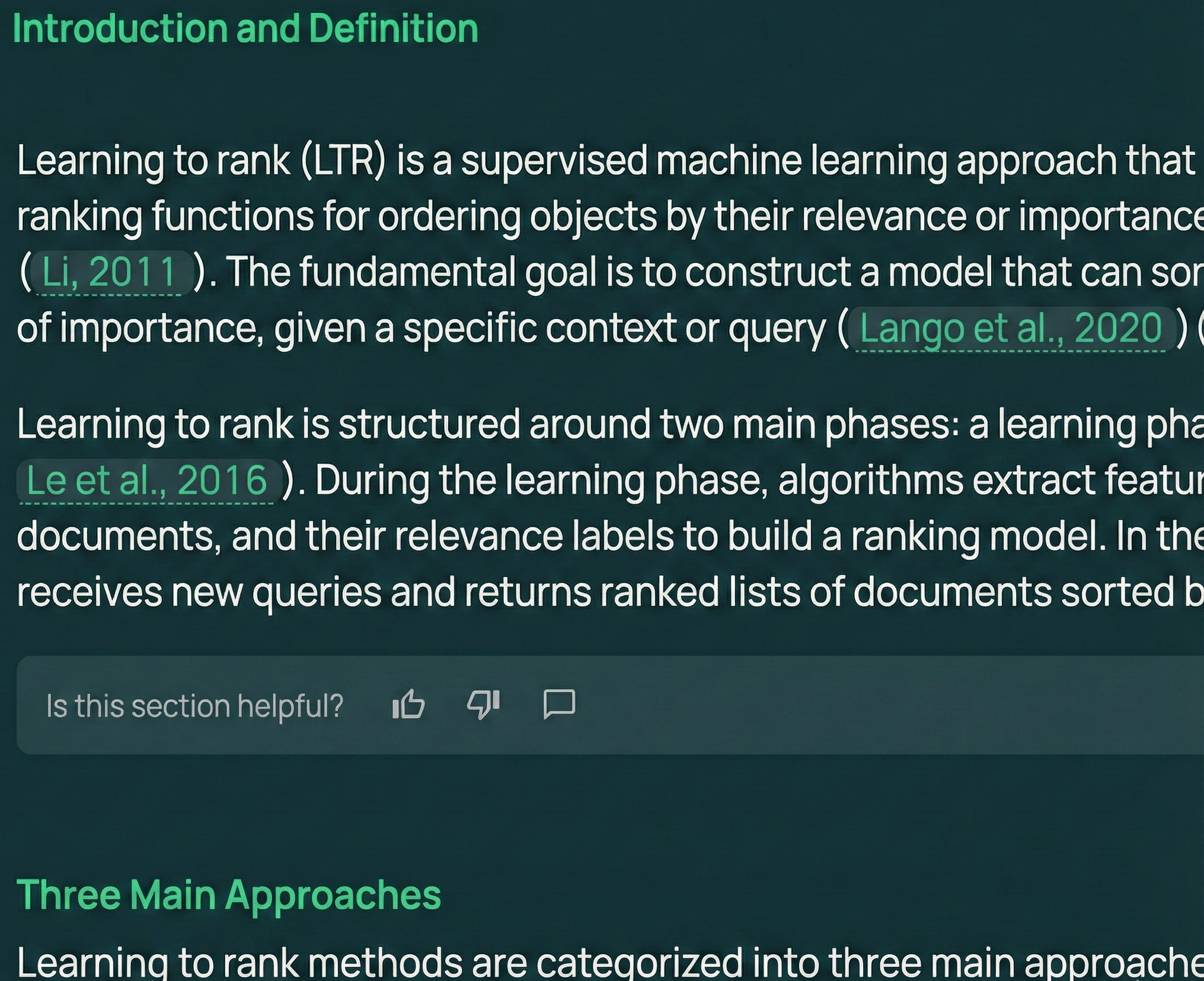}
    \caption{\sqa{} interface. A report with collapsible sections and
      inline citations.}
    \label{fig:sqa-mockup}
  \end{subfigure}
  \caption{Screenshots of the two \sys{} interfaces used in our
    study \citep{singh2025ai2scholarqaorganized,paperfinder2025}.}
  \label{fig:mockups}
\end{figure*}

\textbf{Retrieval and synthesis.} Given a natural-language query, the
system retrieves candidate papers from a scholarly corpus, re-ranks
them, and generates outputs that ground claims in retrieved papers via
inline citations. \pf{} (main interface panel shown in
\Cref{fig:pf-mockup}) presents a chat-like interface returning a
ranked list of papers linked to their Semantic Scholar page, each with a brief generated summary of its
relevance to the query. Users can click to see evidence from the paper
supporting its relevance. \sqa{} (interface shown in
\Cref{fig:sqa-mockup}) is a single-turn literature summary tool producing a
multi-section report: each section has a title, a one-sentence TL;DR
(visible when collapsed), an expandable body with inline citations,
and feedback controls. Citations open evidence cards with a link to the paper page
and excerpts that support the surrounding claim. We refer to the response generated by each query as a \emph{report}: for \sqa{}, this is the multi-section literature summary; for \pf{}, this is the ranked list of papers with generated summaries.

\textbf{Data collected and released.}
We release \dataset, a dataset of anonymized, opt-in user interactions from \sqa and \pf containing 258,935 queries and 432,059 clickstream interactions (February--August 2025). To limit personally identifiable information (PII) risks, we release only hashed report identifiers and drop queries with LLM-detected PII (less than 1\%). Our analysis uses internal pseudonymous user identifiers to compute cohorts and retention; these identifiers are not included in the released dataset to reduce re-identification risk.  The full dataset schema is in Appendix~\ref{app:schema}.


%% file: methods.tex
\section{Analysis design and methodology}
\label{sec:methods}

\noindent\textbf{Analysis pipeline.}
We analyze user behavior through a pipeline combining preprocessing,
LLM-based query labeling, and statistical modeling. Our preprocessing
filters bots, identifies sessions, and removes PII. We label 30,000 single-turn  queries across multiple aspects (intent, phrasing, criteria, field of
study) using GPT-4.1 with structured decoding. 
All statistical tests conducted are two-sided t-tests ($\alpha=0.05$).
We also fit binomial logistic regression models predicting click-through, controlling for false discovery with the Benjamini-Hochberg
procedure over all estimated $p$-values. Full details are in \Cref{app:methods}.

\noindent\textbf{Query taxonomy.}
Traditional IR taxonomies are less appropriate for AI research
assistants like \sys; as users issue complex natural language queries
requiring multi-step reasoning and constraint satisfaction rather than
keyword queries. Therefore, we introduce a new taxonomy with non-mutually
exclusive labels for query intent, phrasing style, and criteria.  We
build our taxonomy through an iterative human-and-LLM process:
starting from manual inspection, we have an LLM (Gemini-2.5-pro)
propose additional labels, then manually consolidate until
convergence. This yields three \textbf{query aspects} (intent,
phrasing style, and criteria) with non-mutually exclusive labels. See
\Cref{app:labels} for full definitions.  We
identify 16 query \textbf{intents} (e.g., \texttt{Broad Topic
  Exploration}, \texttt{Causal and Relational Inquiry}; Table \ref{tab:intent-examples}), 7
\textbf{phrasing styles} (e.g., \texttt{Keyword-style Query},
\texttt{Natural Language Question}; Table \ref{tab:phrasing-examples}), 6 \textbf{criteria} types (e.g.,
\texttt{Temporal Constraints}, \texttt{Methodology-Specific
  Criteria}; Table \ref{tab:criteria-examples}), and 28 fields of study (e.g., \texttt{Biology},
\texttt{Electrical Engineering}, \texttt{Law and Legal Studies}; see \Cref{app:fields} for the full list).

\noindent\textbf{User actions and success metrics.}
We study four primary actions in our clickstream dataset: \emph{\sTwo link clicks} (navigating from the report to a paper page on semanticscholar.org), \emph{section expansions} (revealing the section contents in \sqa), \emph{evidence clicks} (viewing inline citation support), and \emph{feedback} (thumbs up/down). From these we derive: \emph{click-through rate (CTR)}, the fraction of reports with at least one link click; \emph{churn rate}, the fraction of users with no subsequent query; and \emph{return rate}, the fraction of users who return after their initial visit.
Following prior work establishing implicit behaviors as reliable satisfaction indicators~\citep{joachims2005accurately, fox2005evaluating, kim2014modeling}, we use click-through rate (CTR) as our primary success surrogate, as \sTwo{} link clicks strongly predict user returns (\Cref{app:ctr-validation}). Explicit thumbs feedback is too sparse (fewer than 2\% of reports) and less predictive of return than link clicks.

\noindent\textbf{User experience stages.}
To study how behavior evolves with experience, we define three
progression stages based on each user's cumulative query count at the
time of each query: the \emph{\singleQueryDesc} stage (a user's first
query), the \emph{\inexperiencedDesc} stage (queries 2 through 10),
and the \emph{\experiencedDesc} stage (queries beyond the 10th).
These are not separate groups; we track the same users over time as
they progress through these stages, enabling us to observe
within-user behavioral changes as familiarity with \sys grows. About
40\% of users initiate at least 2 queries (the second queries usually
coming within a few hours of the first) while less than 10\% of users
initiate 10 or more queries.


%% file: results.tex
\section{Results}
\label{sec:results}
\noindent
Our analysis reveals that users of both \pf and \sqa initiate queries differently than traditional search engines and engage with results as persistent artifacts.

\noindent\textbf{Query intents.} \Cref{tab:intent-examples} illustrates the
diversity of query intents with representative examples. User queries span
a spectrum from traditional retrieval tasks (finding specific papers,
locating citations, or exploring broad topics) to tasks that go well beyond
search. Users ask for methodological guidance, request help interpreting
their own experimental results, seek to identify research gaps, and even
delegate content generation such as drafting full manuscript sections. This
range suggests users approach \sys{} not merely as a search tool but as a
research assistant capable of supporting the full research workflow.
\Cref{fig:intents} presents quantitative distributions of these intents.

\begin{table*}[t]
  \centering
  \small
  \caption{Representative examples of query intents illustrating of user information needs.}
  \label{tab:intent-examples}
  \renewcommand{\arraystretch}{1.2}
  \setlength{\tabcolsep}{3pt}
  \rowcolors{2}{}{lightgray}  
  \begin{tabularx}{\textwidth}{@{}>{\raggedright}p{4.3cm}X@{}}
    \toprule
    \textbf{Intent} & \textbf{Example Query} \\
    \midrule
    Broad Topic Exploration & \textit{GLP-1 and diabetes} \\
    Specific Factual Retrieval & \textit{What are the four core concepts of Rotter's theory?} \\
    Concept Def. \& Exploration & \textit{Summarize the concept of ``Technoimagination'' [...] by Vilém Flusser} \\
    Comparative Analysis & \textit{Describe the trade-offs between HBr and Cl2 plasma gases for reactive ion etching of polysilicon} \\
    Causal \& Relational Inquiry & \textit{Relation between nighttime digital device use and sleep quality [...]} \\
    Method. \& Proced. Guidelines & \textit{How often should I collect mosquitoes for dengue surveillance?} \\
    Tool \& Resource Discussion & \textit{Are there any tools to count the quality or semantic content of citations?} \\
    Research Gap Analysis & \textit{Survey on the limitations of
                            classical NLP Evaluation Metrics} \\
    Citation \& Evidence Finding & \textit{Can you assist me get the source: Malnutrition refers to deficiencies... (WHO, 2023)} \\
    Specific Paper Retrieval & \textit{Anderson and Moore's paper on the stability of the Kalman filter} \\
    Ideation & \textit{Give me a cost-efficient way to build Rapid Antigen Tests using low-cost expression systems} \\
    Application Inquiry & \textit{ETA prediction with GPS data from cargo} \\
    Data Interpretation Support & \textit{Why do TarM knockout strains show higher IL-1b responses compared to JE2 WT? Discuss my results} \\
    Content Gen. \& Experiment & \textit{Improve this Materials and Methods section for a journal paper [...]} \\
    Academic Document Drafting & \textit{Write a full Materials and
                              Methods section suitable for submission
                              to a peer-reviewed journal in plant
                              science or environmental science. [...]}\\
    \bottomrule
  \end{tabularx}
\end{table*}

\noindent\textbf{Phrasing styles.} Users employ a wide range of phrasing
styles when querying \sys{} (\Cref{tab:phrasing-examples}). While
\texttt{Keyword-style} queries remain the most common phrasing on both
tools (\Cref{fig:phrasing}), a substantial share of queries adopt styles
that would be ineffective on traditional search engines. \texttt{Natural
Language Questions} account
for a large portion of queries, indicating that users
expect the system to parse full sentences and act on directives.
More notably, \texttt{Complex Contextual Narratives}, where users
paste entire draft paragraphs as context before posing a
question, and \texttt{Multi-part Queries} that specify structured
sub-tasks reflect phrasing behaviors shaped by general-purpose LLMs.

\begin{table*}[t]
  \centering
  \small
  \caption{Representative examples of query phrasing styles observed across \pf and \sqa.}
  \label{tab:phrasing-examples}
  \renewcommand{\arraystretch}{1.2}
  \setlength{\tabcolsep}{3pt}
  \rowcolors{2}{}{lightgray}  
  \begin{tabularx}{\textwidth}{@{}>{\raggedright}p{3cm}X@{}}
    \toprule
    \textbf{Phrasing Style} & \textbf{Example Query} \\
    \midrule
    Keyword-style & \textit{piracetam efficacy} \\
    Natural Language Q. & \textit{How are emerging digital technologies reshaping sustainable development outcomes?} \\
    Explicit Instruction & \textit{Write an essay as a doctor of literary discourse analysis[...]} \\
    \makecell[tl]{Complex Context\\Narrative} & \textit{When training large language
                              models, it's essential to develop
                              benchmarks that assess true
                              problem-solving ability, not just
                              factual recall. [\textit{detailed discussion about knowing Berlin's train stations vs.\ planning a novel journey}]
                              [C]onduct a literature review on efforts to curate evaluations that explicitly test synthesis rather than memorization.} \\
    Multi-part Query & \textit{9. Identify and summarise key international instruments (e.g., UN Charter, Budapest Convention, UNSC Resolutions). Discuss their relevance and application [...]} \\
    Boolean/Logical Ops & \textit{(``Family Centered'' OR ``Family Centred'') AND (challeng* OR implantat*)} \\
    \makecell[tl]{Citation/Format\\Specification} & \textit{Use Harvard-style citation to apply in-text citations cite with author's name and publication date, page number to Write `` 1,000-word written reflection [...]} \\
    \bottomrule
  \end{tabularx}
\end{table*}

\noindent\textbf{Search constraints.} Beyond phrasing, many queries
include explicit search constraints that go beyond what traditional
search interfaces support (\Cref{tab:criteria-examples} shows users
applying facets via text descriptions of their search criteria that
traditional search engines cannot handle). As shown in
\Cref{tab:criteria}, \texttt{Methodology-Specific} criteria are by far the
most common constraint on both \sys{} and \sTwo{} (42\% and 29\% of
queries, respectively), reflecting users' desire to filter results by
experimental design or analytical approach. \texttt{Publication Quality}
filters (11\% on \sys{} vs.\ 3\% on \sTwo{}) and \texttt{Temporal
Constraints} (5\% vs.\ 2\%) are also more prevalent on \sys{}. Rarer but
notable are \texttt{Citation/Impact-Based} criteria ($<$1\% on \sys{})
such as requests for highly cited papers or journals above a given impact
factor, and \texttt{Data/Resource Availability} constraints where users
seek studies with publicly available datasets or code.

\begin{table*}[t]
  \centering
  \small
  \caption{Representative examples of search constraints (criteria) specified in user queries across \pf and \sqa.}
  \label{tab:criteria-examples}
  \renewcommand{\arraystretch}{1.2}
  \setlength{\tabcolsep}{3pt}
  \rowcolors{2}{}{lightgray}  
  \begin{tabularx}{\textwidth}{@{}>{\raggedright}p{3cm}X@{}}  
    \toprule
    \textbf{Search Constraint} & \textbf{Example Query} \\
    \midrule
    Methodology-Specific & \textit{federated learning where the authors had coined the new terms for defining client types like homogenous clients, selfish clients...} \\
    Publication Type/Quality & \textit{Please find scientific references only from highly reputable and scopus-indexed international journals...} \\
    Temporal Constraints & \textit{Pharmacological Activities of Ganoderma applanatum. Use journals from 2020-2025} \\
    Metadata-Based & \textit{Papers from SIG CHI that have use cases with an ``adaptive task challenge''} \\
    Cit./Impact-Based & \textit{What is the most cited paper of Orit Hazzan ?} \\
    Data/Resource Availability & \textit{search for documents that provide
                           data of the amount of comic books are
                           purchased among different classes in the
                               United States.} \\
    \bottomrule
  \end{tabularx}
\end{table*}

\subsection{RQ1: How do users query LLM-based information retrieval and synthesis systems, and how does it differ from traditional search?}
\label{sec:rq1}
\begin{figure}[t]
  \centering
  \begin{minipage}[t]{0.48\linewidth}
    \vspace{0pt}
    \centering
    \small
    \captionof{table}{Query complexity and length comparison between \pf, \sqa, and \sTwo
      (mean values with 95\,\% CI). Since \sTwo is a traditional search engine we
      expect that its queries are simpler than those
      performed on \sys (\pf and \sqa). Query length is shown separately below the bar
      as it is not a notion of complexity.}
    \begin{tabular}{@{}lrrr@{}}
      \toprule
      \textbf{Metric} & \textbf{\pf} & \textbf{\sqa} & \textbf{\sTwo} \\
      \midrule
      Constraints & \ci{0.60}{0.05}{0.05} & \ci{0.82}{0.04}{0.08} & \ci{0.15}{0.02}{0.02} \\
      Entities & \ci{4.00}{0.20}{0.15} & \ci{5.14}{0.10}{0.42} & \ci{2.25}{0.05}{0.04} \\
      Relations & \ci{2.17}{0.08}{0.06} & \ci{2.68}{0.02}{0.18} & \ci{1.20}{0.04}{0.04} \\
      \midrule
      Length & \ci{17.04}{2.51}{2.38} & \ci{36.96}{6.82}{9.02} & \ci{5.35}{0.18}{0.17} \\
      \bottomrule
    \end{tabular}
    \label{tab:query-metrics}
  \end{minipage}
  \hfill
  \begin{minipage}[t]{0.48\linewidth}
    \vspace{0pt}
    \centering
    \includegraphics[width=\linewidth]{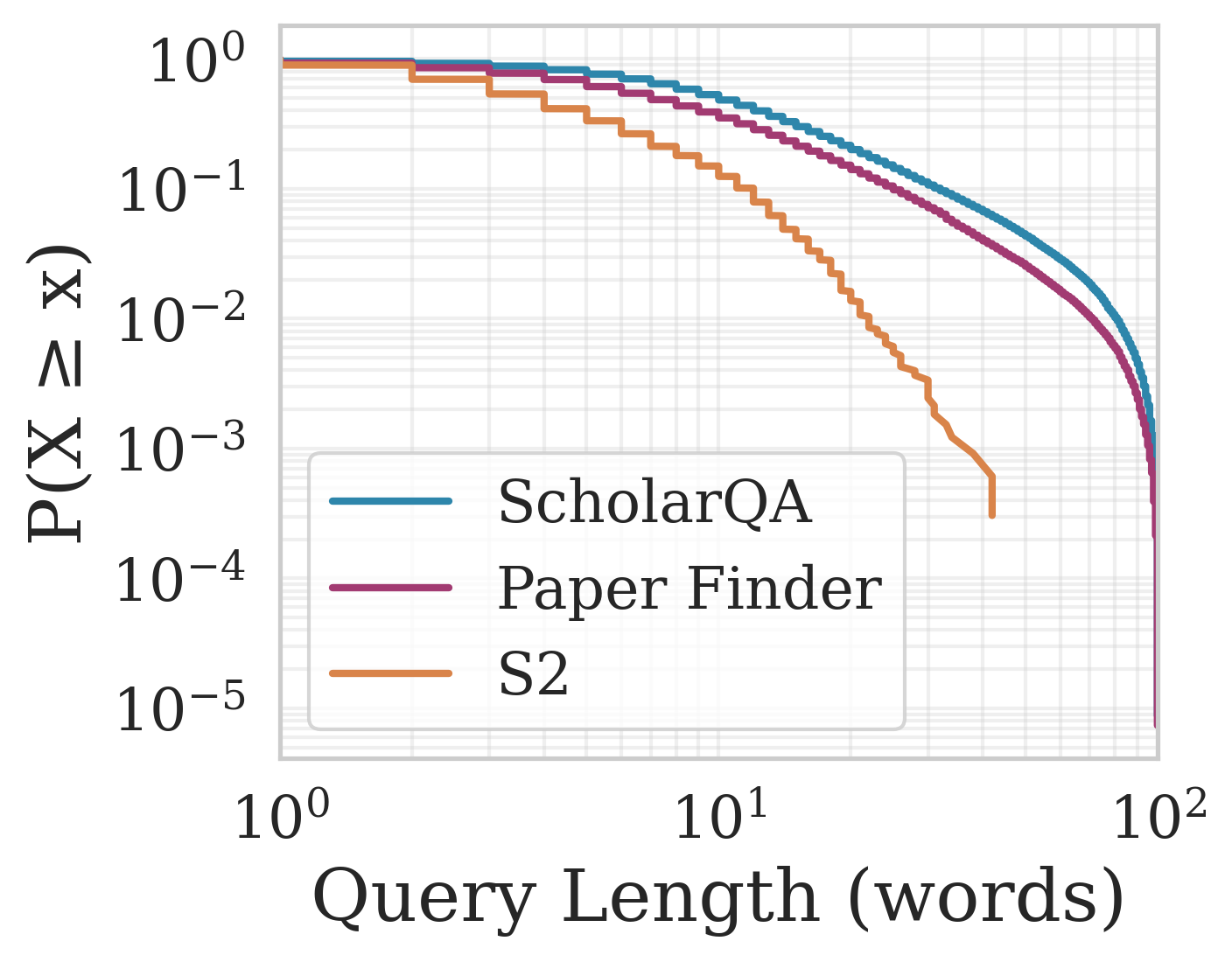}
    \caption{Query length distribution for \pf and \sqa showing heavy-tailed behavior.}
    \label{fig:query-length-power-law}
  \end{minipage}
\end{figure}

We analyze session duration, query complexity, and query types across tools and over time. Users initiate much longer and more complex queries than on the \sTwo{} baseline (Table~\ref{tab:query-metrics}). Query complexity on \sTwo{} itself has also increased over time, possibly due to broader AI adoption raising user expectations.

\noindent\textbf{Session duration and usage patterns.}
Users typically initiate 1-2 queries per session with a median
session duration of 4 minutes for \pf and 8 minutes for
\sqa.\footnote{Note that the median response time for \pf and \sqa are
  34 seconds and 129 seconds, respectively.} The
median number of sessions per user and the median number of queries
per user is 2 for both \pf and \sqa. Like
many IR tools, there is a wide spread of usage level, with both
session duration and number of sessions per user having a long
tail. The shorter duration of \pf sessions is to be expected since one
of the main actions performed on \pf (clicking on a \sTwo link)
navigates the user away from \pf (although in a new tab) whereas most
of the content presented by \sqa is consumed in-situ. Across both tools, usage is dominated by repeat users: twice as many of our users have
initiated multiple queries as have initiated only a single query, and more
than half of those repeat users have initiated more than 15 queries.

\noindent\textbf{Temporal changes in query behavior.}
As previously noted, we saw \sTwo{} queries' complexity also increase over the years. Comparing \sTwo{} queries between 2022 and 2025, we find that the fraction of queries with at least 1
constraint rose from 7\,\% to 10\,\%, queries with at least 1 relation
grew from 65\,\% to 78\,\%, and average query length increased from
4.8 to over 6 words. This suggests users increasingly expect search
systems to handle more complex queries, likely shaped by exposure to
AI-powered tools.

\noindent\textbf{Query intent and phrasing distribution.}
While query length and structural complexity capture surface-level properties, they do not reveal what users are trying to accomplish. We therefore examine
the distribution of query intents, phrasing styles, fields of study,
and criteria specified in the
query (with labels identified using an LLM).

\begin{figure}[t]
  \centering
  \begin{subfigure}[t]{0.48\linewidth}
    \centering
    \includegraphics[width=\linewidth]{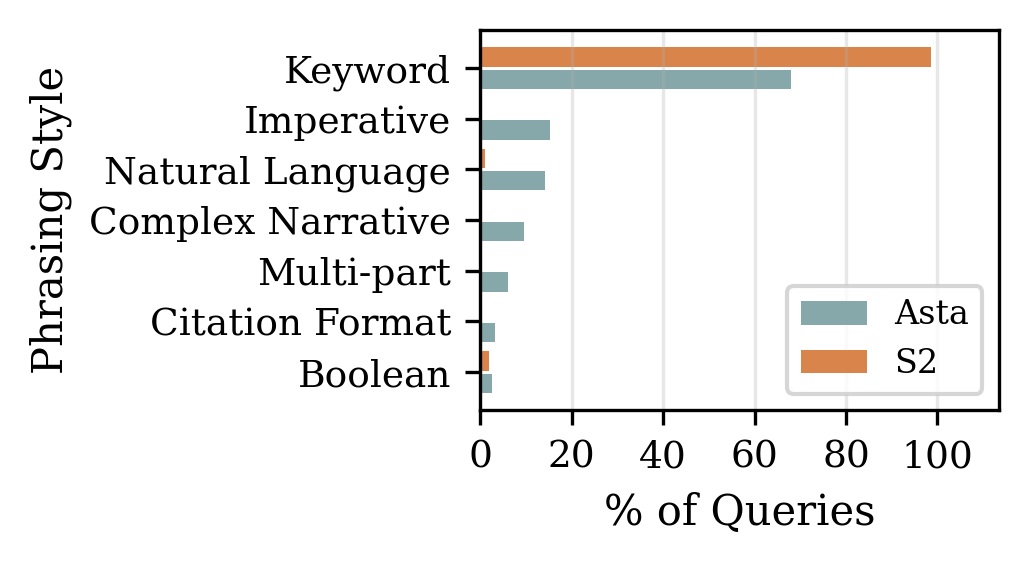}
    \caption{Phrasing Style}
    \label{fig:phrasing}
  \end{subfigure}
  \hfill
  \begin{subfigure}[t]{0.48\linewidth}
    \centering
    \includegraphics[width=\linewidth]{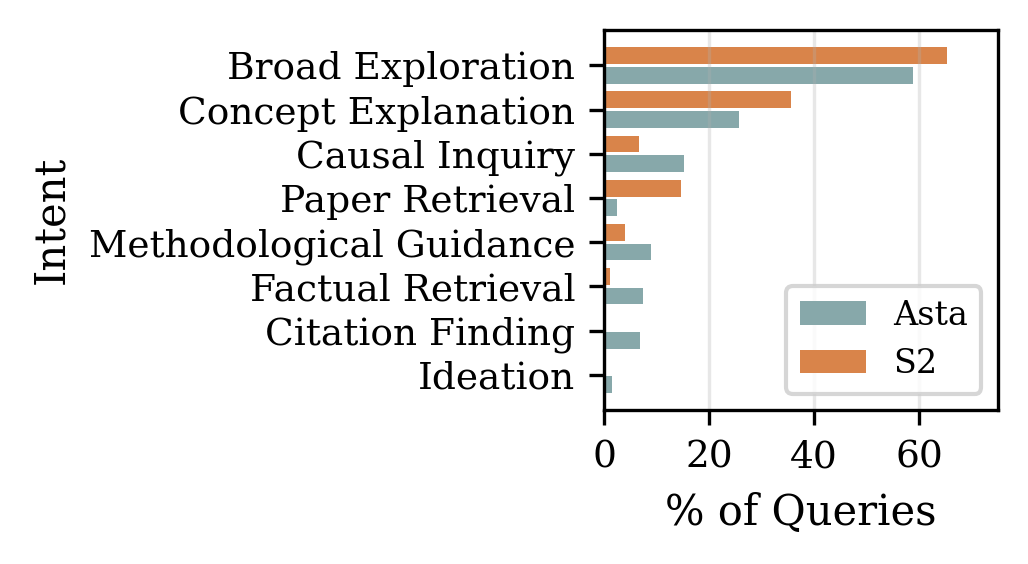}
    \caption{Intent}
    \label{fig:intents}
  \end{subfigure}
  \caption{Query phrasing styles and intents (\% of queries). Keyword queries dominate the distribution. \sys users initiate more natural language queries than \sTwo users. Most queries are broad exploration or concept explanation; \sTwo is skewed towards broad and specific paper retrieval queries. See \Cref{tab:intent-fractions} for complete intent distribution, and \Cref{tab:criteria} for criteria distribution.}
  \label{fig:query-distribution}
\end{figure}

\Cref{fig:query-distribution} shows query distribution across these
various aspects and reveals that relatively simple queries are the
most common: \texttt{Broad Topic Exploration} and
\texttt{Keyword-style} queries dominate the query distribution. Fewer
than half of all queries specify some explicit search criteria
constraint (see \Cref{tab:criteria} for the full list of rates), the
most common being criteria regarding paper methodology (see \Cref{tab:criteria-examples}). In contrast, queries to \sTwo are skewed
more heavily toward simpler queries: nearly all (98\%) of queries to
\sTwo are keyword-style queries. The query intents that are relatively
more common in \sTwo
are also simpler and reflect the informational and
navigational queries of \citet{broder2002taxonomy} from traditional IR
systems: broad queries and specific paper retrieval queries.

We also observe variation in query intents across scientific fields. For example,
computer science queries are the most likely to be \texttt{Ideation} queries while history queries are the least
likely (see \Cref{fig:field-intents} in the appendix for the full
breakdown). Moreover, the prevalence of \texttt{Ideation} queries on \sys{} compared to \sTwo{}
(\Cref{fig:intents}) suggests that users are asking AI-powered tools
to take over tasks they would previously have done themselves, not
just retrieving relevant papers, but directly generating ideas and
solutions.

\noindent\textbf{User expectations.}
The distribution of query intents shows that users expect \sys{} to
perform diverse tasks beyond search, including experimental design and
identifying unexplored research directions. Yet despite these
expectations, keyword-style queries and broad topic exploration
dominate, a pattern that persists even among experienced users
(\Cref{sec:cohort}), suggesting functional fixedness. Multi-part queries or \texttt{Complex Contextual Narrative} queries often include long passages copy-pasted from drafts followed by a question, taking advantage of LLMs' capability of handling long texts.
Beyond typical queries, users employ diverse strategies (template-filling, explicit prompting, collaborative writing workflows, and refinding queries) that reveal expectations shaped by general-purpose LLMs (see \Cref{tab:interesting-query-patterns} in the appendix for examples). These patterns suggest users expect \sys{} to function as a collaborative research partner with capabilities similar to general-purpose chatbots.
Users also increasingly use abstract concepts rather than specific jargon (e.g., ``why some language models behave unpredictably when trained further'' vs.\ ``BERT fine-tuning instability''); 66\% of \pf queries include abstract concepts vs.\ 38\% in \sTwo{} (\Cref{app:abstractiveness}).

\noindent\textbf{Learning effects.}
\label{sec:cohort}
Tracking the same users across the three experience stages defined in
\Cref{sec:methods}, we find that users learn to issue more targeted
queries over time. As shown in \Cref{tab:progression-queries},
\texttt{Broad Topic Exploration} drops from 61.2\% in the
\singleQueryDesc stage to 53.5\% in the \experiencedDesc stage, while
other query intents become more common. This suggests that as users
gain more experience with the system they learn to initiate more
complex and challenging queries beyond broad topic exploration queries.

\begin{table}[t]
  \centering
  \small
  \caption{Query label by experience stage, tracking the same user population over time (percentages with 95\,\% confidence intervals).}
  \label{tab:progression-queries}
  \begin{tabular}{@{}lrrr@{}}
    \toprule
    Query Label & \singleQueryDesc & \inexperiencedDesc & \experiencedDesc \\
    \midrule
    Broad Topic Exploration & \ci{61.23}{1.38}{1.38} & \ci{55.63}{0.98}{0.98} & \ci{53.48}{0.99}{0.99} \\
    Causal and Relational Inquiry & \ci{15.78}{1.03}{1.03} & \ci{15.97}{0.72}{0.72} & \ci{17.06}{0.75}{0.75} \\
    Citation \& Evidence Finding & \ci{6.25}{0.69}{0.69} & \ci{8.65}{0.55}{0.55} & \ci{9.65}{0.59}{0.59} \\
    Methodology-Specific Criteria & \ci{41.84}{1.40}{1.40} & \ci{45.50}{0.98}{0.98} & \ci{47.17}{0.99}{0.99} \\
    \bottomrule
  \end{tabular}
\end{table}

\noindent\textbf{Near-duplicate queries and report revisitation.}
Report revisitation is common: 50.5\% of \sqa users and 42.1\% of \pf users revisit previous reports, substantially more than near-duplicate query submission (18.8\% and 14.8\%, respectively). This suggests users treat \sys{} results as persistent artifacts, reference material that they can consult multiple times, rather than ephemeral search results. Near-duplicate queries occur on shorter timescales (median $<$16 minutes) than revisits (median 4--6 hours), with most showing slight refinements such as format instructions or language preferences (see \Cref{tab:duplicate-vs-revisit,tab:duplicate-examples} in the appendix).

\subsection{RQ2: How do users engage with the presented content?}
\label{sec:rq2}

Our analysis of engagement patterns reveals that users learn to extract additional value from these systems as they gain experience: \sqa{}
users discover that they can verify claims by examining the cited
works (the rate of clicking on inline evidence increases by 27\%
between a user's first query and their 4th), while experienced \pf{} users increasingly consume information
directly from the result list without clicking through to papers (link
clicks drop by 24\% over the same period), as
the interface provides sufficient context.

\noindent\textbf{\sqa-specific actions.}
\sqa presents reports as expandable sections with TL;DR summaries. Analysis reveals diverse, non-linear reading behaviors: users skip the introduction 43\% of the time, and over half of reports (52.4\%) involve non-consecutive section expansions. While sequential reading dominates, users frequently skip sections, navigate backwards, and return to the introduction from later sections (see \Cref{app:sqa-navigation} for detailed transition analysis and \Cref{fig:reading} for the Sankey diagram). These patterns suggest that the section-oriented display helps users efficiently identify and access just the portions of information most relevant to their needs.

\noindent\textbf{Churn due to latency and errors.}
Since AI-powered tools have higher latency and error rates than traditional search, we
examined how response time and errors affect user retention. \pf
typically returns results within 30 seconds while \sqa takes around 2
minutes to generate full reports. Users tolerate this difference: \sqa
churn remains stable (close to 11\,\%) for response times under 5
minutes, whereas \pf churn increases by 10\,\% relative if responses
exceed 1 minute, suggesting that users expect \pf to behave like traditional
search but accept longer waits for \sqa's synthesized reports. In
contrast, catastrophic errors severely impact retention: first-time
users who encounter an error have only a 10\,\% chance of returning,
compared to 53\,\% for users with an initial successful experience.


%% file: related.tex
\section{Related work}
\label{sec:related}

Understanding user behavior has long been central to information
retrieval (IR) research. Classic web search work proposes query intent
taxonomies including
Broder's influential three-way split into informational,
navigational, and transactional intents~\citep{broder2002taxonomy} and
subsequent refinements~\citep{rose2004understanding,jansen2008determining,li2010classifying,hashemi2020intent}.
These taxonomies have primarily been developed for keyword-based web
search over general-purpose corpora. We similarly develop and apply a
taxonomy of user intents, but focus on scientific research tasks and
LLM-based assistants.

Recent work has begun to characterize how people query and adapt to
LLM-based information access systems. \citet{liu2025trapped} examine ``functional
fixedness'' in LLM-enabled chat search, showing how users' prior
experience with search engines, virtual assistants, and LLMs
constrains their initial prompt styles and adaptation strategies, and
proposing a typology of user intents in chat
search. \citet{shah2024taxonomy} introduce an
LLM-plus-human-in-the-loop pipeline to generate and apply user intent
taxonomies from large-scale Bing search and chat logs, uncovering
distinct intent distributions between traditional search and AI-driven
chat and demonstrating how such taxonomies support log analysis at
scale. Complementing these log-based
perspectives, \citet{wang2024ux} develop a taxonomy of seven high-level user
intents for general LLM interactions and, through a survey of 411
users, reveal heterogeneous usage patterns, satisfaction levels, and
concerns across intents. \citet{kim2024followup} focus
specifically on conversational search, deriving a taxonomy of 18
follow-up query patterns and using an LLM-powered classifier on
real-world logs to relate different follow-up behaviors to user
satisfaction signals. Together, these studies
show that LLM-based systems elicit rich behaviors that are strongly mediated by
users' mental models and prior tool use, but they primarily
investigate general-purpose search or everyday LLM services rather
than domain-specific scientific research tools.

Another line of work examines how LLM-powered search affects
engagement with information and the diversity of content that users
ultimately see. \citet{spatharioti2025effects} compare LLM-powered search with more traditional search
conditions in decision-making tasks, measuring differences in speed,
accuracy, and overreliance. \citet{kaiser2025newera} conduct a large-scale study of user behavior and preferences
during practical search tasks, contrasting generative-AI search with
traditional search engines and documenting differences in
how people explore and prefer results across
interfaces.

While some work exists characterizing usage patterns with LLM-based
systems, these studies release only final analyses rather than
underlying interaction data.  Anthropic~\citep{tamkin2024clioprivacypreservinginsightsrealworld}
and OpenAI~\citep{chatterji2025how} released limited descriptions of
how people use their chat products, including LLM-derived taxonomies
of user intents.  \citet{yang2025adoption} present a study of AI agent adoption
analyzing usage from millions of Perplexity
users.  OpenRouter's State of AI report
analyzes task distributions, model preferences, and retention
patterns~\citep{openrouter2025stateofai}.  Other white papers address
the \textit{economic} impacts of AI
tools~\citep{handa2025economictasksperformedai,appel2025anthropiceconomicindexreport}.
None of these publicly release interaction data. The
LMSYS~\citep{zheng2024lmsyschat1mlargescalerealworldllm},
WildChat~\citep{zhao2024wildchat1mchatgptinteraction}, and Open
Assistant~\citep{köpf2023openassistantconversationsdemocratizing}
datasets do release user conversation text, but include only basic
metadata and are broad-domain rather
than specific to systems targeting researchers.

%% file: discussion.tex
\section{Discussion}
\label{sec:discussion}
Our analysis reveals that users pose longer, more complex queries to
AI-powered tools than to traditional search-powered tools. Users treat
results as persistent artifacts, and experienced users engage more
deeply with content than other users. However, these patterns must be interpreted in light of differential query success across query types, which we discuss under ``limitations and generalizability'' below.

\noindent\textbf{Implications for tool and interface design.}
The behavioral patterns  we observe suggest several directions for the design of
future AI-powered scientific research tools.

\textit{Query formulation support.}
The observation that users often discover unmet requirements only after
seeing initial results (evidenced by near-duplicate queries that add
format instructions or language preferences) suggests value in
clarifying user intent before executing long-running queries.
Users' feedback often resembles follow-up instructions (e.g., ``add a
historical component''), indicating expectations for iterative
refinement through conversation. The high rate of report revisitation
(42--50\,\% of users) and users' treatment of results as persistent
artifacts suggest that generated content may need mechanisms for
staying current as new literature appears.

\textit{Content navigation and consumption.}
The non-linear reading patterns we observe in \sqa{}, including
frequent skipping of introductions (43\,\% of reports) and revisiting
of specific sections, motivate designs that foreground section-level
navigation, TL;DR-style summaries, and user control over the ordering
and granularity of generated content, rather than assuming strictly
sequential consumption.

\textit{Reliability and latency tolerance.}
Users tolerate higher latency for report-generation tools (\sqa{})
than for search-oriented tools (\pf{}), but are highly sensitive to
catastrophic errors during their initial query. This error-sensitivity suggests that graceful
degradation and clear error recovery paths may be especially important for
first impressions.

\textit{Addressing underserved query types.}
Users expect \sys{} to
function as a general-purpose agentic system. Content generation
queries, temporal constraints, data resource requests, and citation
format specifications all show lower satisfaction (see \Cref{sec:limitations} for the underlying odds ratios), and the behavioral
patterns underlying these failures are striking. Users submit
template-filling queries, explicit prompting instructions, and
collaborative writing workflows (\Cref{tab:interesting-query-patterns}),
all of which presuppose a general-purpose conversational agent rather
than a task-specific retrieval tool.
At the same time, \texttt{Complex Contextual Narrative} queries, where
users provide extensive context by pasting from drafts or describing
their research situation in detail, are among the most successful
query types on \sqa{} (OR\,=\,1.47), suggesting that when the
interface allows users to supply rich context, the system can
effectively leverage it.

\noindent\textbf{Limitations and generalizability.}
\label{sec:limitations}
Our behavioral findings study one representative system, \sys{}, and the findings may be skewed toward query types that this system happens to address well.  To help quantify this potential bias, we estimate which queries \sys{} handles more effectively today by fitting logistic regression models predicting click-through rate (CTR) from query attributes for each tool with user features as controls (see \Cref{app:ctr-validation} for a justification of using CTR as a success surrogate). For \pf{}, Citation/Evidence Finding (OR\,=\,1.17) and Broad Topic Exploration (OR\,=\,1.12) queries have higher click odds, while Content Generation and Expansion (OR\,=\,0.47), Data Resource Availability (OR\,=\,0.61), and Temporal Constraint (OR\,=\,0.82) queries have substantially lower click-through odds. For \sqa{}, Concept Definition and Explanation (OR\,=\,1.29) and Complex Contextual Narrative (OR\,=\,1.47) queries have higher click odds, while Citation Format Specification (OR\,=\,0.62) queries have lower click odds. See \Cref{app:query-success} for the full analysis, odds ratio tables, and coefficient visualizations.

\section{Conclusions and Future Work}

We have presented the Asta Interaction Dataset, interaction logs from \pf{} and \sqa{}, two LLM-based research assistants deployed within \sys{}. While both tools follow standard design patterns for this class of system, our findings may not generalize to tools with substantially different retrieval scope, interaction modality, or optimization objectives. We view our released dataset, taxonomy, and behavioral analysis as a starting point for cross-system comparisons and more targeted experiments.

Future work will examine follow-up queries and user journeys over
time, tracking how users refine their queries within and across sessions and how
their mental models of system capabilities evolve with experience. We
also plan to investigate cross-tool usage patterns, characterizing how
users move between \pf{} and \sqa{} within research workflows and what
triggers transitions between search-oriented and report-oriented
tools.


%% file: appendix_methods.tex
\appendix
\crefalias{section}{appendix}

\makeatletter
\renewcommand{\@seccntformat}[1]{%
  \ifnum\pdfstrcmp{#1}{section}=0
    Appendix~%
  \fi
  \csname the#1\endcsname\quad
}
\makeatother

\section{Methods Details}
\label{app:methods}

\subsection{Query Analysis Pipeline}
\label{app:query-analysis-pipeline}
Our preprocessing pipeline includes bot and canned query filtering, session identification, action debouncing, and PII removal. We limit analysis to a stable deployment period and focus on single-query behavior. Session boundaries are set based on a 45-minute UI action timeout period. Action debouncing is performed for page revisits: we only consider a page to be revisited if it occurred after 5 minutes from the initial visit. To preserve privacy, we remove any queries flagged by an LLM as possibly containing PII. 
Since we have left the study of followup query behavior for future study, we focus primarily on user behavior with respect to a single query at a time. Specifically, we restrict \pf
analysis to the first query in each \pf conversation as neither \sqa nor \sTwo
supported chat-style followup queries during the data collection period.
We have removed any notion of user ID from our data release.

We use LLMs throughout our analysis pipeline: for query labeling (intent, phrasing, criteria, field of study), complexity assessment, duplicate detection, feedback classification, and response quality evaluation. All prompts are provided in \cref{app:prompts}, with label definitions and few-shot examples in \cref{app:labels}.
Using the taxonomy we label a random subset of 30,000 queries. For each query aspect, we prompt GPT-4.1 with the query text, possible labels, descriptions, and in-context examples, using structured decoding to ensure valid outputs. Labels are non-mutually exclusive; a query might receive both \texttt{Broad Topic Exploration} and \texttt{Methodological Guidance} for intent.

\subsection{Comparison Between Tools}
\label{app:comparison-methodology}
Since \pf is naturally a multiturn chat experience whereas \sTwo
and \sqa neither support multi-turn conversations or contextualized
followup queries, we limit our analysis in this study to the first
query initiated in a \pf conversation. This may end up skewing the query
distribution. Followup queries are left to future work.

\subsection{Statistical Analysis}
\label{app:statistical-analysis}
Our study explores many aspects of user behavior; we use frequentist
hypothesis testing to identify statistically significant findings and
focus on those with the large effect sizes. All reported differences between groups (e.g., rates across tools, user segments, or query types) are statistically significant at $\alpha=0.05$ using two-sided t-tests. We report 95\% confidence intervals (Wilson CIs for rates, bootstrap for unbounded values), showing the larger bound as a single $\pm$ value.
To identify query characteristics associated with successful result engagement, we fit binomial logistic regression models predicting click-through rate (CTR) for each tool. In this context, a \texttt{click} is a binary outcome indicating whether a user clicked on at least one Semantic Scholar paper result following their query.
Query level covariates were derived from LLM-based multilabel
classifications of each query, expanded into binary indicators
(including those from the query taxonomy). To control for
heterogeneity in usage patterns with the tools between users, we included user history statistics as covariates: the number of previous queries issued by the user, cumulative prior clicks, and an empirical Bayes-smoothed estimate of the user's historical click rate.
Separate models were fit for \sqa and \pf (since each tool exposes a different interface) using maximum likelihood estimation with a logit link ($n=30,000$ for each model). To control the false discovery rate when testing the significance of individual coefficients, we applied the Benjamini-Hochberg procedure across all p-values from both models.


%% file: appendix_analysis.tex
\section{Abstractiveness Analysis}
\label{app:abstractiveness}

The transition to natural language queries, together with users' expectations from modern AI systems, increased the use of abstract intents, meaning that users shifted from relying on jargon terms to expressing their information needs more abstractly. For example, a query like ``BERT fine-tuning instability'' increasingly appears in forms such as ``why some language models behave unpredictably when trained further'', where the specific technical term is replaced by an abstract description of the underlying intent.

We measured the abstractiveness of queries by classifying abstract concepts versus jargon terms using an LLM. We found that 38\% of \sTwo{} queries include at least one abstract concept, compared with 66\% in \pf{}. The median number of abstract concepts per query is 0 in \sTwo{} and 1 in \pf{}, reinforcing the same trend. We also observe a positive correlation between query length and the number of abstract concepts it contains (Pearson $r = 0.519$). Since \pf{} queries tend to be longer, this suggests that users now express their intents through more elaborate, abstract descriptions, while \sTwo{} queries were shorter and more densely packed with jargon. This substantial gap indicates that query complexity increased not only due to the higher rate of entities and relations, but also because modern queries deviate more from the scientific jargon anchored in the documents themselves.

\section{SQA Section Navigation Analysis}
\label{app:sqa-navigation}

\begin{figure}[tb]
  \centering
  \includegraphics[width=0.95\linewidth]{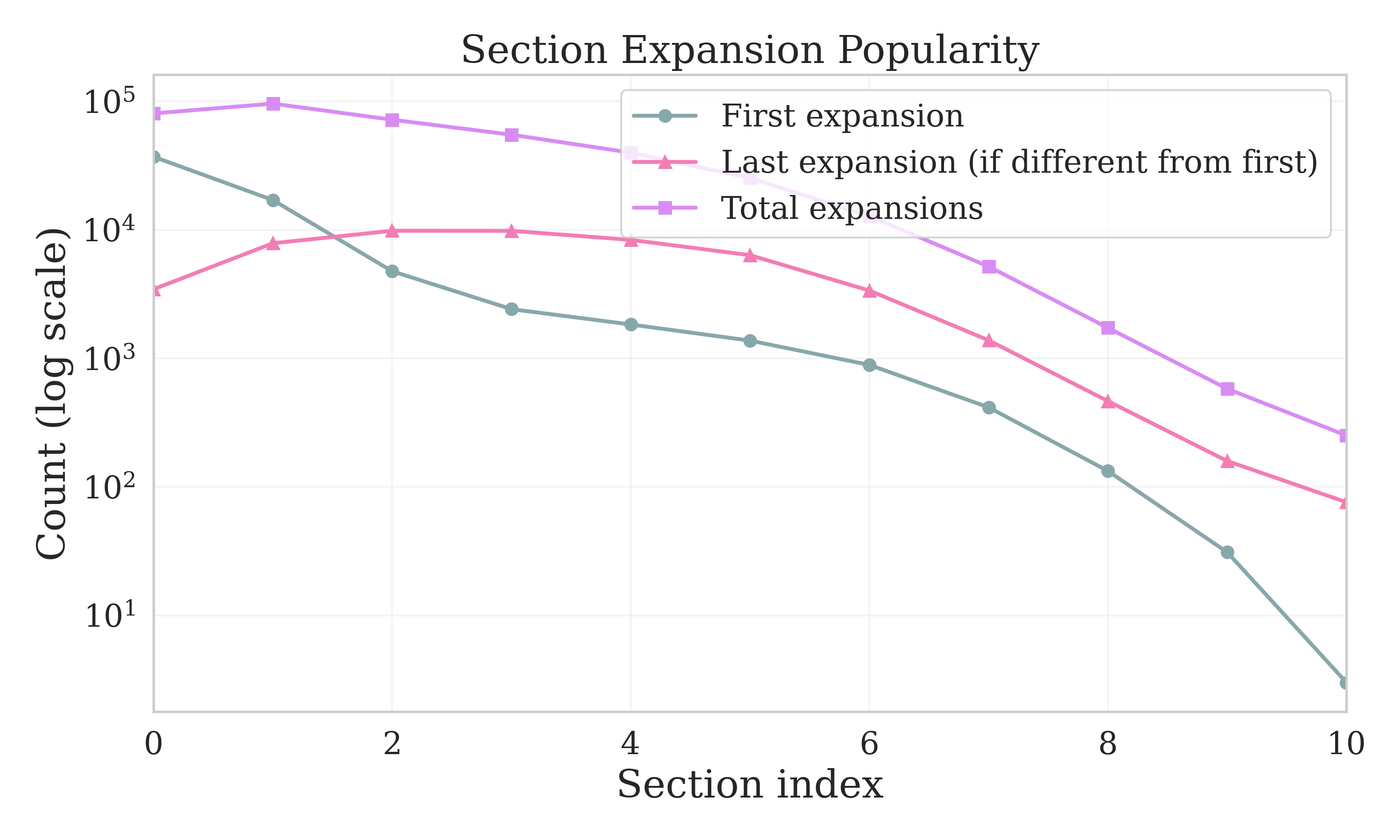}
  \caption{Section expansion distribution (on a log scale) showing which sections users expand first in \sqa responses. Section index 1 has the largest number of expansions. Users tend to start on section 0 or 1 and end on a section between 2--4.}
  \label{fig:sqa-section-expansion}
\end{figure}

\Cref{fig:sqa-section-expansion} shows the distribution of section expansions. Users often start from the first section (index 0), and the last section expanded in a report is typically between index 1 and 4. Given that the first section is almost always an introduction to the topic, it is notable to see users frequently starting with the second section instead. We do still see the expected position bias towards the first sections of the report, with the total number of expansions on each section index over all reports decreasing with index number after index 1; the tendency to skip the introduction appears to be strong enough that index 1 is the most commonly expanded section.

\begin{figure}[tb]
  \centering
  \includegraphics[width=0.95\linewidth]{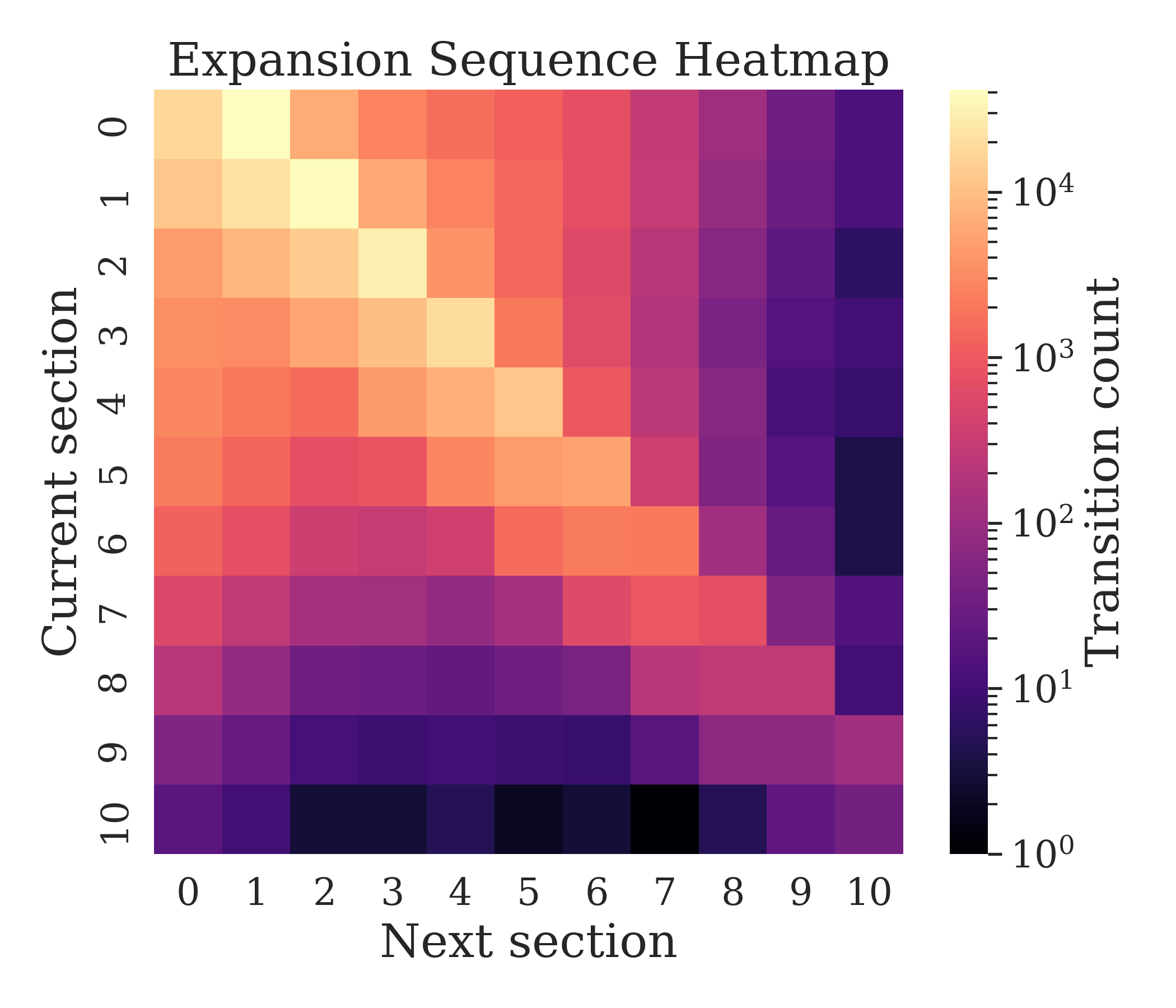}
  \caption{Section transition heatmap showing in-order vs out-of-order reading patterns in \sqa. Sequential expansion is the dominant behavior, but there is notable backward traversal behavior as well as return to 0 (the introduction) behavior. Users also close and then reopen a section (presumably to read the TL;DR which is only visible when a section is collapsed).}
  \label{fig:sqa-section-transitions}
\end{figure}

To better understand user navigation behavior and the importance of sequential generation of the report content, we examined user navigation between sections. \Cref{fig:sqa-section-transitions} shows the section transition counts, specifically how many times a user expanded section $j$ after having just expanded section $i$. Notably, the upper off-diagonal is bright, indicating sequential expansions, but users also exhibit the behavior of closing and reopening a section (presumably to read the TL;DR which is only visible when the section is collapsed). There is also notable section skipping behavior and a bright lower off-diagonal indicating sequential expansion backwards. Also notable is seeing users often return to section 0 (the introduction section) regardless of the section they are currently on. Overall, users primarily move sequentially through the report with some skipping behavior and backwards navigation.

\section{CTR Validation as Success Metric}
\label{app:ctr-validation}

While thumbs up/down feedback would directly indicate satisfaction, it
is sparse (less than 2\% of reports; see
\Cref{fig:overall-action-distribution}) and available only for
self-selecting users. We therefore use link clicks as our success
proxy: they occur on at least 15\% of \pf reports and correlate more
strongly with repeat usage (users who click links are more likely to
return than those who provide thumbs up feedback, see
\Cref{fig:action-return-rates} for full analysis).

We also considered using \pf evidence clicks but chose link clicks to
maintain parity between tools. LLM-assessed quality of the report
generated by \sqa has a substantial correlation with success
metrics. Reports assessed as high quality have an average CTR of
5.5\,\% compared to 3.8\,\% for reports assessed as low quality, a
relative increase of 44\,\%. Similarly, the return rate of reports
assessed as high quality is 62.3\,\% versus 54.7\,\% for low-quality
reports. The relative size of these differences validates the
importance of response quality, though the absolute change suggests
that quality, at least as estimated by an LLM judge, is only one
factor affecting user behavior.

%% file: appendix_results.tex
\section{Additional figures and tables}
\label{app:figs}

\begin{table*}[t]
  \centering
  \footnotesize
  \caption{Query intent distribution (\% of queries) with random examples.}
  \label{tab:intent-fractions}
  \begin{tabularx}{\textwidth}{@{}lrX@{}}
    \toprule
    \textbf{Intent} & \textbf{\%} & \textbf{Example} \\
    \midrule
    \multicolumn{3}{@{}l}{\textbf{\sqa}} \\
    \addlinespace[2pt]
    Broad Topic Expl. & 51.6 & \textit{What are the types of approaches to employer branding?} \\
    Concept Def.\ \& Expl. & 28.2 & \textit{Digital Banking BI Maturity Models} \\
    Causal \& Rel.\ Inquiry & 19.1 & \textit{why evaluation as a construct of metacognition is very important for writing competence to emerge?} \\
    Specific Factual Retr. & 12.6 & \textit{what do say paper of sharma2019 and kamga 2016 about resistace in chemotherapy in AML IN NOTCH pathway} \\
    Method.\ \& Proc.\ Guid. & 9.1 & \textit{design non-stationnary
                                     wavelets trough [...] bezout's equation} \\
    Comparative Analysis & 7.3 & \textit{conduct an expansive analysis
                                 on the methods for and efficiency of
                                 using acoustic analysis along with
                                 machine learning for forest fire
                                 detection [...]} \\
    Acad.\ Doc.\ Drafting & 6.2 & \textit{Write a [...] section for a master's thesis titled ``The Role of Digital Twin Technology in Improving Construction Project Delivery''} \\
    Cit./Evid.\ Finding & 5.7 & \textit{[Chinese: I want to prove that higher MFN brainwave amplitude may be due to higher cognitive load or cognitive conflict]} \\
    Focused Acad.\ Synth. & 5.6 & \textit{SIR Model Study (papers from 2020 to 2024).} \\
    Res.\ Gap \& Limit.\ Anal. & 5.2 & \textit{are there studies that look into robustness to reward noise on domains other than math for RLVR?} \\
    App.\ Inquiry & 4.2 & \textit{how can a new leader on a unit reduce staff burnout} \\
    Tool \& Res.\ Discovery & 2.3 & \textit{Video to urdf format} \\
    Complex Cross-Paper Synth. & 2.1 & \textit{start with Gestational diabetes mellitus and reach to impaired insulin signalling in placenta and mediated role of inostols} \\
    Prob.\ Solving \& Ideation & 1.7 & \textit{What research is done into generating scientific ideas, specifically predicting the next paper based on the previous citations using LLMs} \\
    Data Interp.\ \& Anal. & 1.1 & \textit{FTO was electrochemically etched using HCl (2M) and zinc acetate (0.03M) at different etching time... the etching rate decrease exponentially with time... why?} \\
    Content Gen.\ \& Exp. & 1.1 & \textit{[Arabic: rewrite and restructure methodology section for doctoral dissertation on IT and event management]} \\
    Specific Paper Retr. & 0.7 & \textit{Jayasinghe, A 2014, ``Broadlands power project will kill Kitulgala's white water rafting''} \\
    \midrule
    \multicolumn{3}{@{}l}{\textbf{\pf}} \\
    \addlinespace[2pt]
    Broad Topic Expl. & 65.0 & \textit{Enzymatic electrolysis} \\
    Concept Def.\ \& Expl. & 23.5 & \textit{iot framework} \\
    Causal \& Rel.\ Inquiry & 12.0 & \textit{[lit review for underrepresented learners in rural Nigerian communities]} \\
    Method.\ \& Proc.\ Guid. & 8.9 & \textit{[PhD proposal guidance for autonomous satellite control to avoid space debris using AI]} \\
    Cit./Evid.\ Finding & 7.7 & \textit{Could you help me find a few CNS-level articles that discuss the knockout of DRP1 in cells and then its overexpression?} \\
    App.\ Inquiry & 5.0 & \textit{[Chinese: femtosecond laser + sapphire substrate, manipulating nanosheets]} \\
    Focused Acad.\ Synth. & 4.5 & \textit{Influence of Contaminated Ammonium Nitrate on Detonation Behaviour of Bulk Emulsion Explosives and Numerical Analysis of Detonation-Induced Damage Zone} \\
    Specific Paper Retr. & 4.1 & \textit{https://onlinelibrary.wiley.com/doi/10.1002/pds.5880} \\
    Comparative Analysis & 3.8 & \textit{[German: digital service platforms + automated assessments, functional \& non-functional requirements]} \\
    Tool \& Res.\ Discovery & 2.9 & \textit{Custom query languages for graphs that [...]} \\
    Specific Factual Retr. & 2.8 & \textit{[Chinese: metastatic TNBC first-line chemotherapy median progression-free survival (mPFS) is only 5-6 months]} \\
    Complex Cross-Paper Synth. & 2.5 & \textit{can you find papers that criticize Kilian and Vigfusson 2011, 2013 work and confirm the presence of asymmetries or nonlinearities in the oil-macroeconomy relationship} \\
    Res.\ Gap \& Limit.\ Anal. & 2.3 & \textit{[systematic review abstract on conversational agent interventions for physical and psychological symptom management]} \\
    Prob.\ Solving \& Ideation & 1.1 & \textit{[Portuguese: write a thesis proposal on spatial geometry with a didactic intervention in high school]} \\
    Data Interp.\ \& Anal. & 0.6 & \textit{[...] all models consistently struggled to differentiate the Moderate Risk group[...] individuals in moderate states exhibit overlapping characteristics with both low and high-risk groups} \\
    Content Gen.\ \& Exp. & 0.6 & \textit{[Portuguese: write and compile thesis on drumstick leaves bioavailability]} \\
    \bottomrule
  \end{tabularx}
\end{table*}

\begin{table}[tb]
  \centering
  \small
  \caption{Query criteria distribution (\% of queries). Methodology-related criteria are most common on \sys.}
  \label{tab:criteria}
  \begin{tabular}{@{}lrr@{}}
    \toprule
    & \textbf{\sys} & \textbf{\sTwo} \\
    \midrule
    Methodology & 42 & 29 \\
    Pub.\ Quality & 11 & 3 \\
    Metadata & 9 & 12 \\
    Temporal & 5 & 2 \\
    Data Avail. & 1 & 0.4 \\
    Cit.\ Impact & 0.8 & $<$0.1 \\
    \bottomrule
  \end{tabular}
\end{table}

\begin{figure}[tb]
  \centering
  \includegraphics[width=0.95\linewidth]{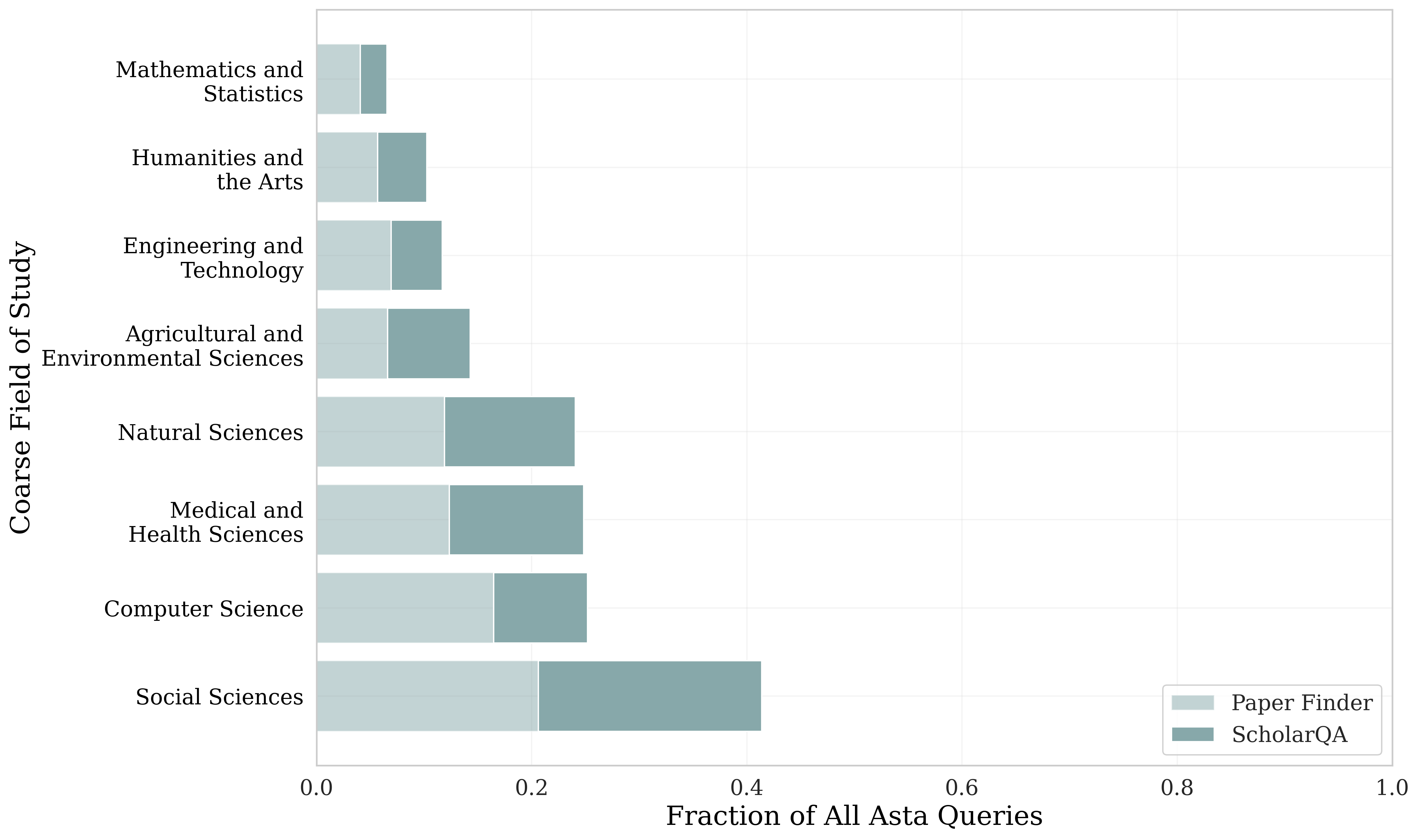}
  \caption{Distribution of coarse-grained fields of study across \pf, \sqa, and \sTwo.}
  \label{fig:fields}
\end{figure}

\begin{figure}[tb]
  \centering
  \includegraphics[width=0.95\linewidth]{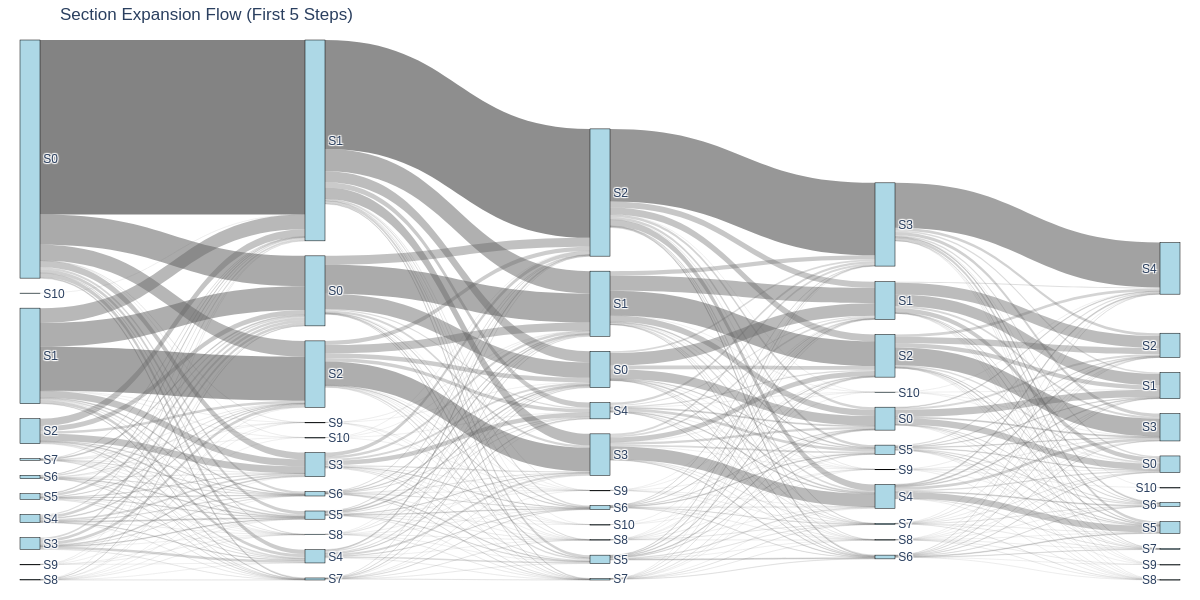}
  \caption{Reading order shown through section flow Sankey diagram, illustrating user navigation patterns through \sqa response sections.}
  \label{fig:reading}
\end{figure}

\begin{table}[h]
\centering
\small
\begin{tabular}{p{0.25\columnwidth}p{0.65\columnwidth}}
\toprule
\textbf{Characteristic} & \textbf{Description} \\
\midrule
Keyword-style Query (phrasing style) & Short, often fragmentary, queries resembling search engine keywords or subject headings. No verbs or complete sentences; typically a list of nouns or concepts separated by spaces or simple punctuation. \\
\addlinespace
Methodology-Specific Criteria (query criteria) & Queries where the user requires specific methods, approaches, analytical techniques, or experimental designs to be present in the search results. This can include demands for computational models, experimental paradigms, or meta-analyses. \\
\bottomrule
\end{tabular}
\caption{Common query characteristics observed in \sys.}
\label{tab:characteristics-examples}
\end{table}

\begin{figure*}[tb]
  \centering
  \includegraphics[width=0.95\linewidth]{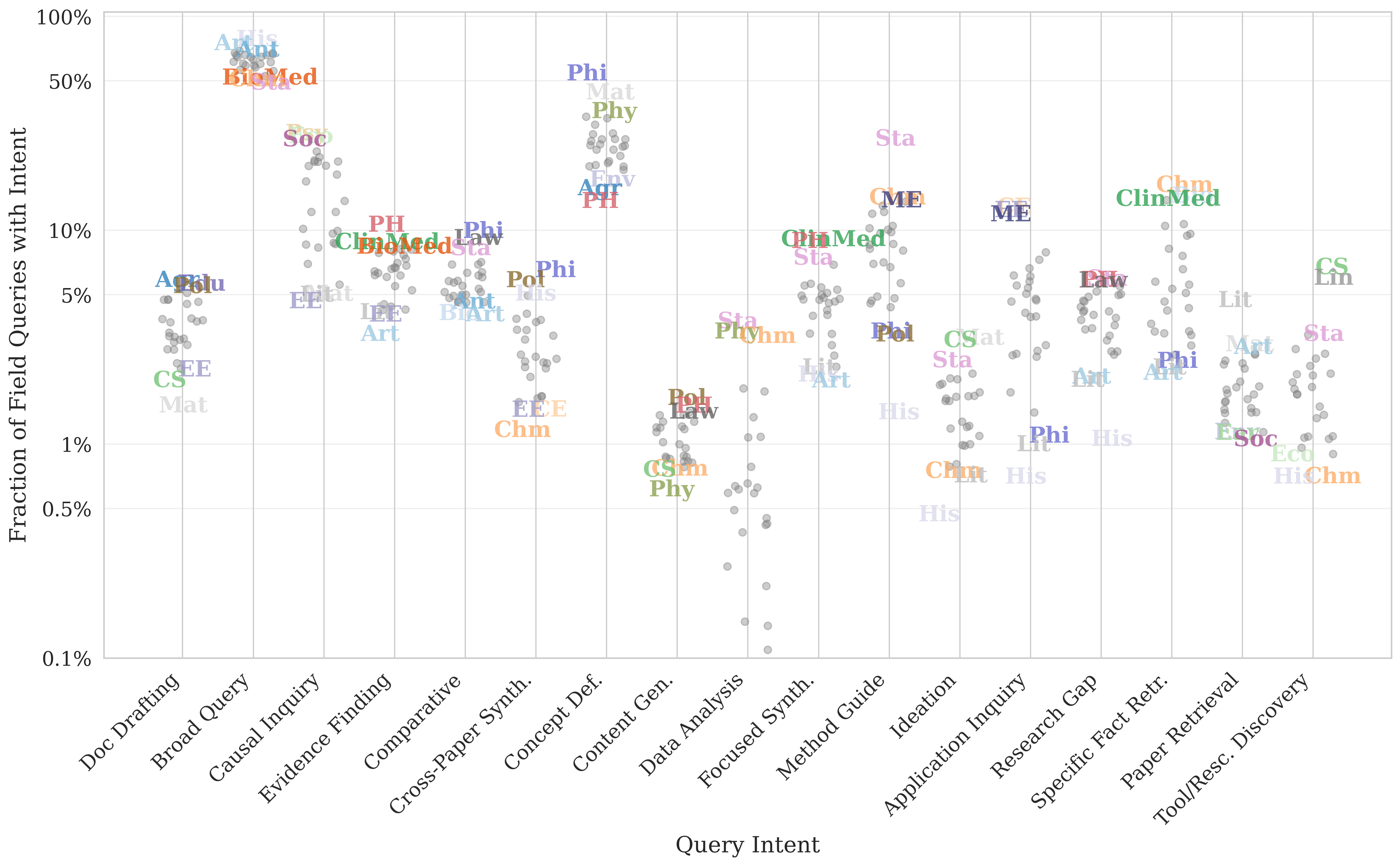}
  \caption{The fraction of queries from each field of study that have
    a given intent on \sys. The distribution reflects the research
    tasks common in the given fields.}
  \label{fig:field-intents}
\end{figure*}

\begin{figure}[tb]
  \centering
  \includegraphics[width=0.95\linewidth]{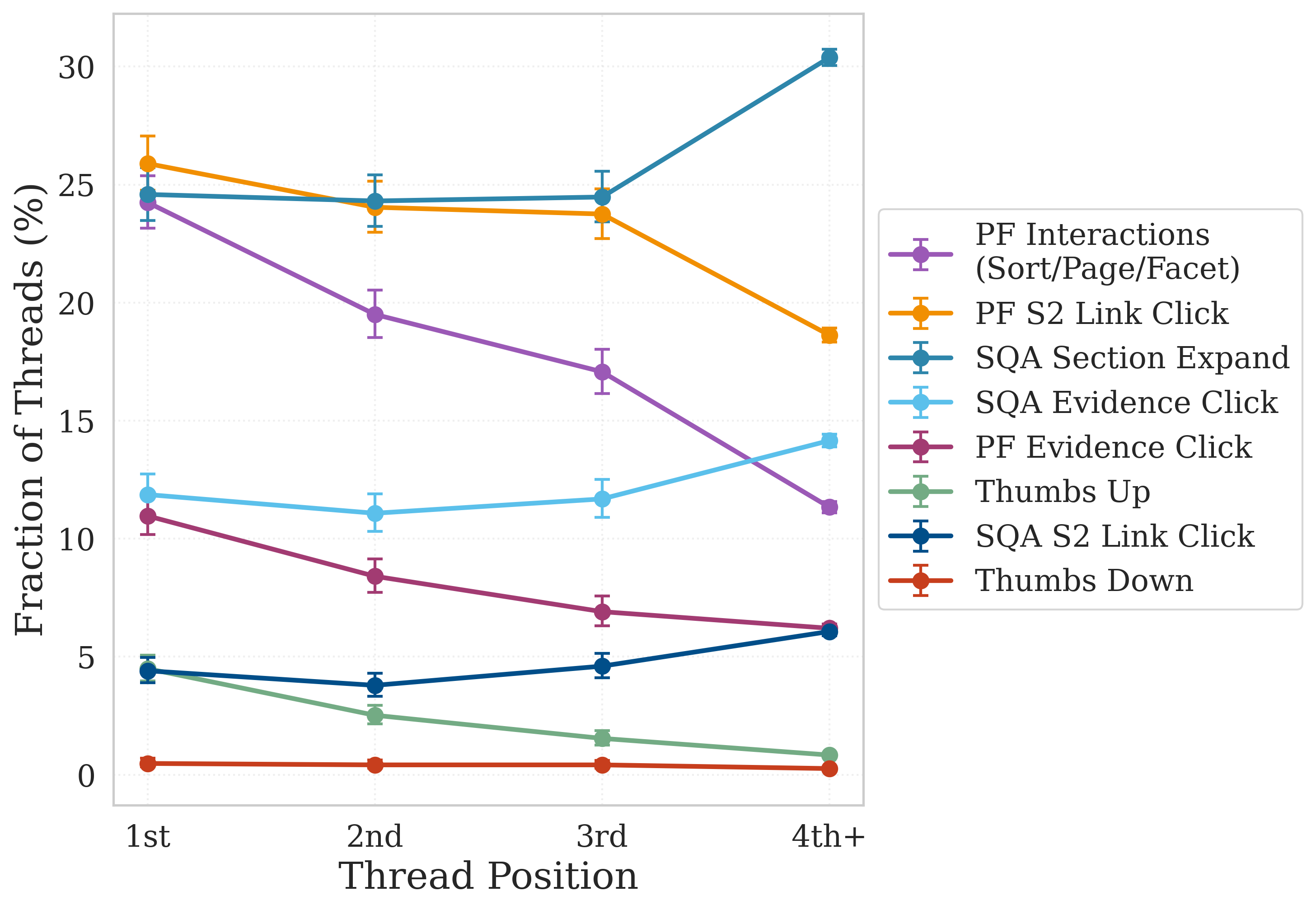}
  \caption{Action engagement trends by query index showing how users
    perform different actions as they gain experience with the
    system. Shaded regions represent 95\,\% confidence intervals. \pf
    reports tend to receive less click engagement over time compared
    to \sqa reports which grow in engagement, the reason likely being
    that \pf results can be consumed passively without interacting
    with the web content at all whereas most of the content generated
    by \sqa can only be accessed after clicking on the web page.}
  \label{fig:action-comparison}
\end{figure}

\begin{figure}[tb]
  \centering
  \includegraphics[width=0.95\linewidth]{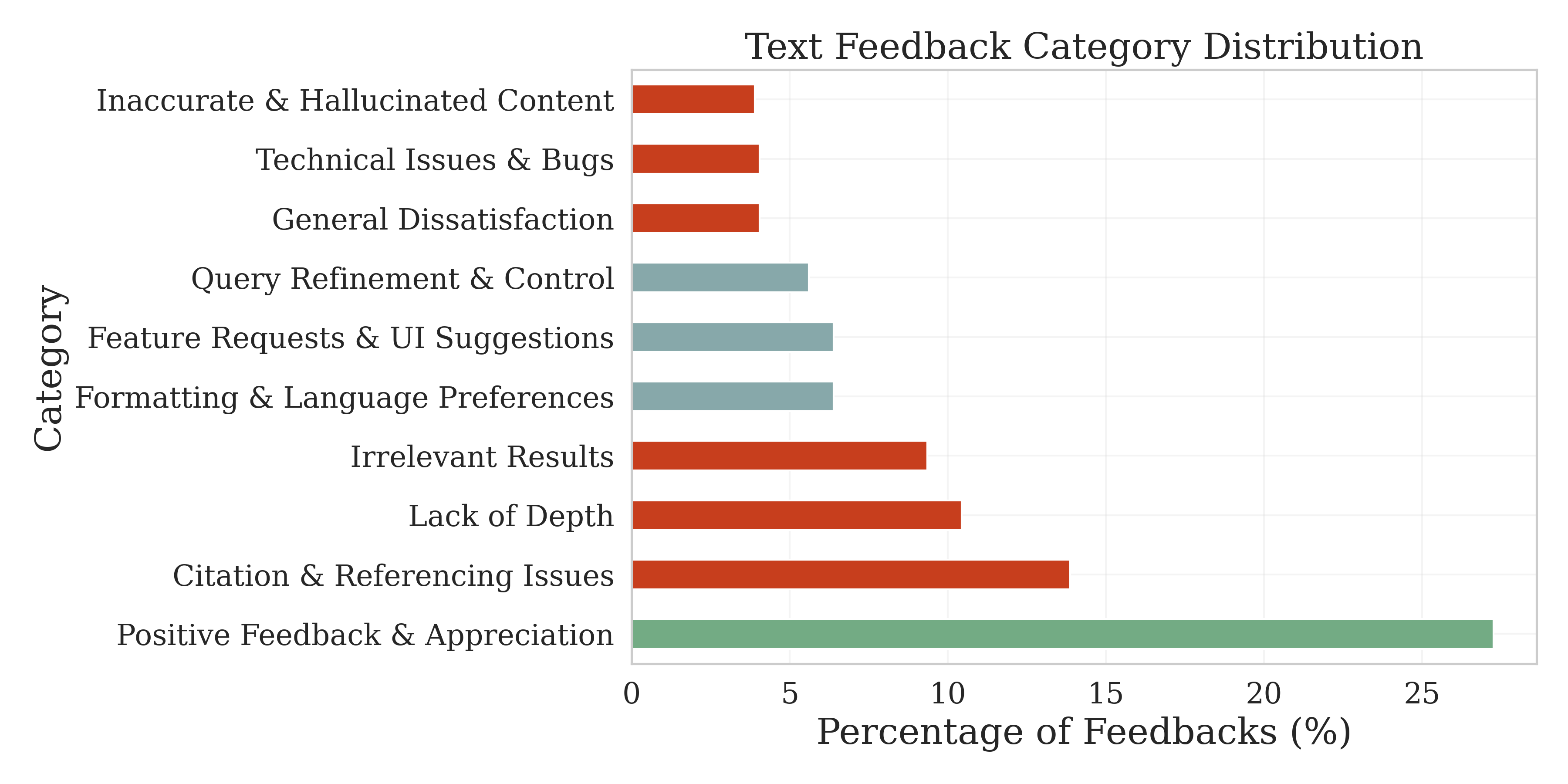}
  \caption{Distribution of text feedback categories reflecting user
    expectations and experiences. The comprehensiveness of the
    response is the most common complaint (\texttt{Lack of Depth} and
    \texttt{Citation \& Referencing Issues}).}
  \label{fig:feedback-categories}
\end{figure}

\begin{table}[t]
  \centering
  \small
  \caption{Example user feedback by category. Some users seem to
    expect the feedback submission to provide a multiturn chat experience.}
  \label{tab:feedback-examples}
  \begin{tabularx}{\linewidth}{lX}
    \toprule
    \textbf{Category} & \textbf{Example Feedback} \\
    \midrule
    Response Quality & develop more the introduction and add more citations (references) \\
    \addlinespace
    Response Quality & this is leaving out some key details from Magalhaes et al 2013 \\
    \addlinespace
    Response Quality & Out of date \\
    \addlinespace
    Query Refinement & Its very good but I need to add a historical component \\
    \addlinespace
    Feature Requests & would be a great place to a diagram \\
    \addlinespace
    Feature Requests & amazing! Coudl you create a PDF \\
    \addlinespace
    Formatting Preferences & too dense. I wanted bullet points \\
    \bottomrule
  \end{tabularx}
\end{table}

\begin{table*}[t]
  \centering
  \small
  \caption{Examples of interesting usage patterns showing how users
    probe the capabilities of LLM-powered research tools. Users go
    beyond basic IR tasks and expect \sys to work as a collaborative
    research assistant.}
  \label{tab:interesting-query-patterns}
  \begin{tabularx}{\textwidth}{llX}
    \toprule
    \textbf{Pattern} & \textbf{Tool} & \textbf{Example Query} \\
    \midrule
    Template Filling & \pf & \textit{fill this tabel with 10 jurnal bellow:...} [table template and citations] \\
    \addlinespace
    Template Filling & \sqa & \textit{for sacubitril find all: "IUPAC Name: CAS Number: Molecular Formula:..."} [15+ fields] \\
    \addlinespace
    Prompting & \sqa & \textit{You are an expert research assistant specializing in computational geosciences and machine learning.} \\
    \addlinespace
    Prompting & \pf & \textit{Find papers...The model **must** be capable of...} \\
    \addlinespace
    Persona Adoption & \sqa & \textit{Think of yourself as experienced
                              professor...Please write me a phd
                              proposal...devour
                              Turnitin detection bots} \\
    \addlinespace
    Collaborative Writing & \sqa & \textit{I'm working on my paper...} [LaTeX section] \textit{add papers from TSE, TOSEM, ICSE} \\
    \addlinespace
    Collaborative Writing & \pf & [Chinese paragraph] \textit{Help me find references...tell me which sentences cite which} \\
    \addlinespace
    Research Lineage & \pf & \textit{What are latest advances in research fields of these three papers?} [3 DOIs] \\
    \addlinespace
    Refinding & \pf & \textit{...paper using BERT that says we
                           cant just look at top-k...which paper says
                           this} \\
    Refinding & \pf & \textit{hey whats the name of the paper that did a study on how people use llms by allowing the public to use their tokens on paid llms...} \\
    \bottomrule
  \end{tabularx}
\end{table*}

\begin{table}[t]
  \centering
  \small
  \caption{Comparison of duplicate query submission versus report
    revisiting behavior.}
  \label{tab:duplicate-vs-revisit}
  \begin{tabular}{@{}lrr@{}}
    \toprule
    \textbf{Metric} & \textbf{\sqa} & \textbf{\pf} \\
    \midrule
    \multicolumn{3}{l}{\textit{User Prevalence}} \\
    Users with duplicate queries & 18.8\,\% & 14.8\,\% \\
    Users who revisited reports & 50.5\,\% & 42.1\,\% \\
    \addlinespace
    \multicolumn{3}{l}{\textit{Temporal Patterns}} \\
    Median time between duplicates & 15.9 min & 5.8 min \\
    Median time between revisits & 3.8 hours & 5.9 hours \\
    \addlinespace
    \multicolumn{3}{l}{\textit{Short-term occurrence ($<$1 hour)}} \\
    Duplicates within 1 hour & 66.7\,\% & 72.9\,\% \\
    Revisits within 1 hour & 30.5\,\% & 23.5\,\% \\
    \addlinespace
    \multicolumn{3}{l}{\textit{Medium-term occurrence ($<$1 day)}} \\
    Duplicates within 1 day & 81.2\,\% & 83.5\,\% \\
    Revisits within 1 day & 72.2\,\% & 71.2\,\% \\
    \bottomrule
  \end{tabular}
\end{table}

\begin{table}[t]
  \centering
  \footnotesize
  \caption{Representative examples of duplicate queries showing exact
    duplicates and incremental refinements. Time gaps indicate minutes
    between submissions.}
  \label{tab:duplicate-examples}
  \begin{tabularx}{\linewidth}{llX}
    \toprule
    \textbf{Tool} & \textbf{Gap} & \textbf{Query Pair} \\
    \midrule
    \sqa & 4.0 min & \textit{First:} History of experiential learning of science \\
    & & \textit{Second:} History of experiential learning of science \\
    \addlinespace
    \sqa & 1.4 min & \textit{First:} postbiotics food \\
    & & \textit{Second:} postbiotics in food industry \\
    \addlinespace
    \sqa & 6.4 min & \textit{First:} [Long query requesting literature review] \\
    & & \textit{Second:} [Same query + ``write this in future tense''] \\
    \addlinespace
    \pf & 4.1 min & \textit{First:} Item-based collaborative filtering recommendation algorithm \\
    & & \textit{Second:} Item-based collaborative filtering recommendation algorithm \\
    \addlinespace
    \pf & 0.9 min & \textit{First:} find me papers that experiment with dropping entire LLM blocks \\
                  & & \textit{Second:} find me papers that have to do with dropping entire blocks from decoder transformer LLMs \\
    \addlinespace
    \pf & 3.4 min & \textit{First:} [Chinese query about gated fusion attention] \\
    & & \textit{Second:} [Same + ``respond in Chinese''] \\
    \bottomrule
  \end{tabularx}
\end{table}

\begin{figure}[htbp]
  \centering
  \includegraphics[width=0.95\linewidth]{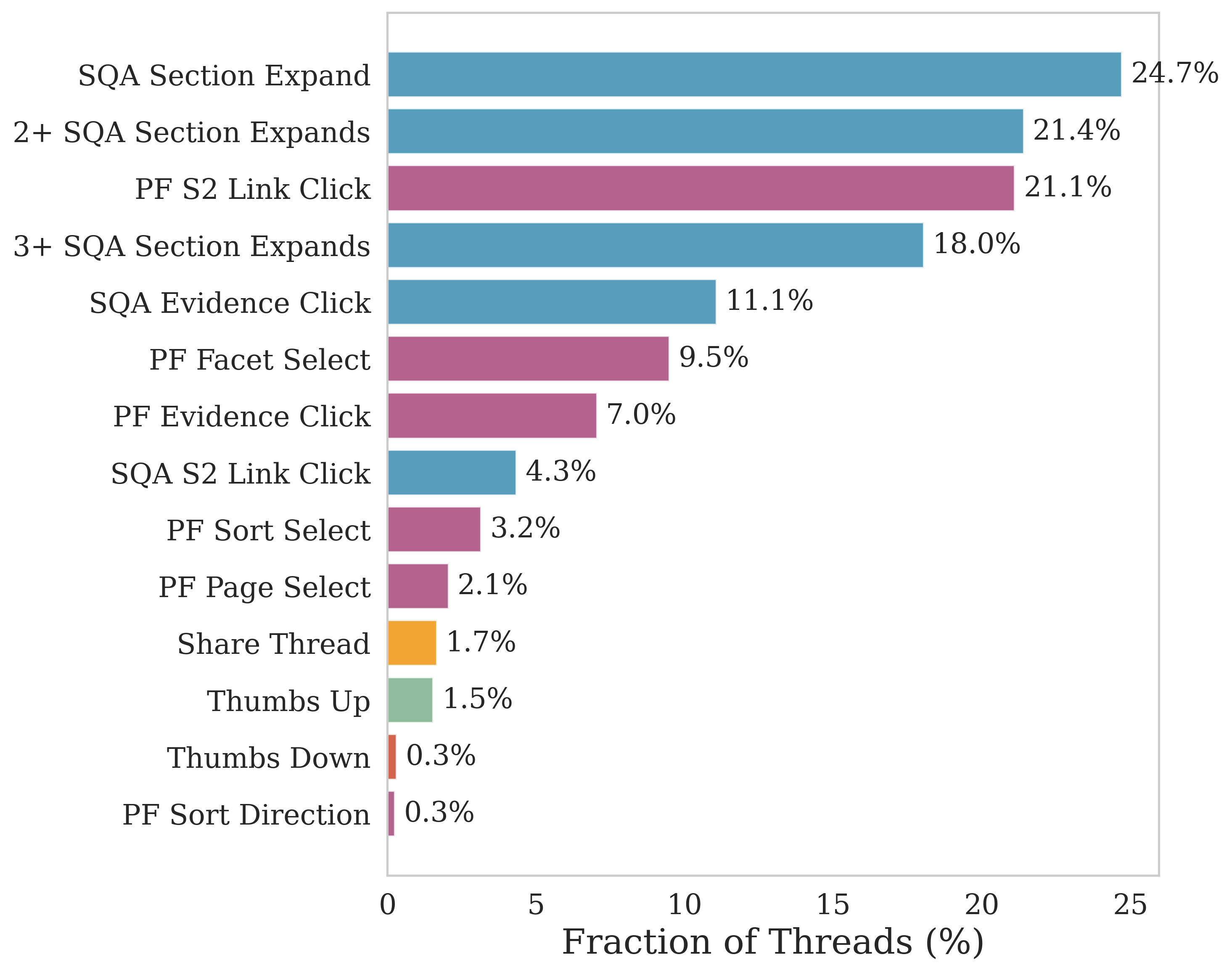}
  \caption{Fraction of reports on which the given action has been
    performed across \pf and \sqa, showing how users engage with
    different features. Section expansion actions are the most common,
    likely because users must click to expand the section text in the
    \sqa response. Note that link clicks are much more common than
    thumbs up/down feedback; below we also show that link clicks are
    also strongly associated with use satisfaction.}
  \label{fig:overall-action-distribution}
\end{figure}

\begin{figure}[htbp]
  \centering
  \includegraphics[width=0.95\linewidth]{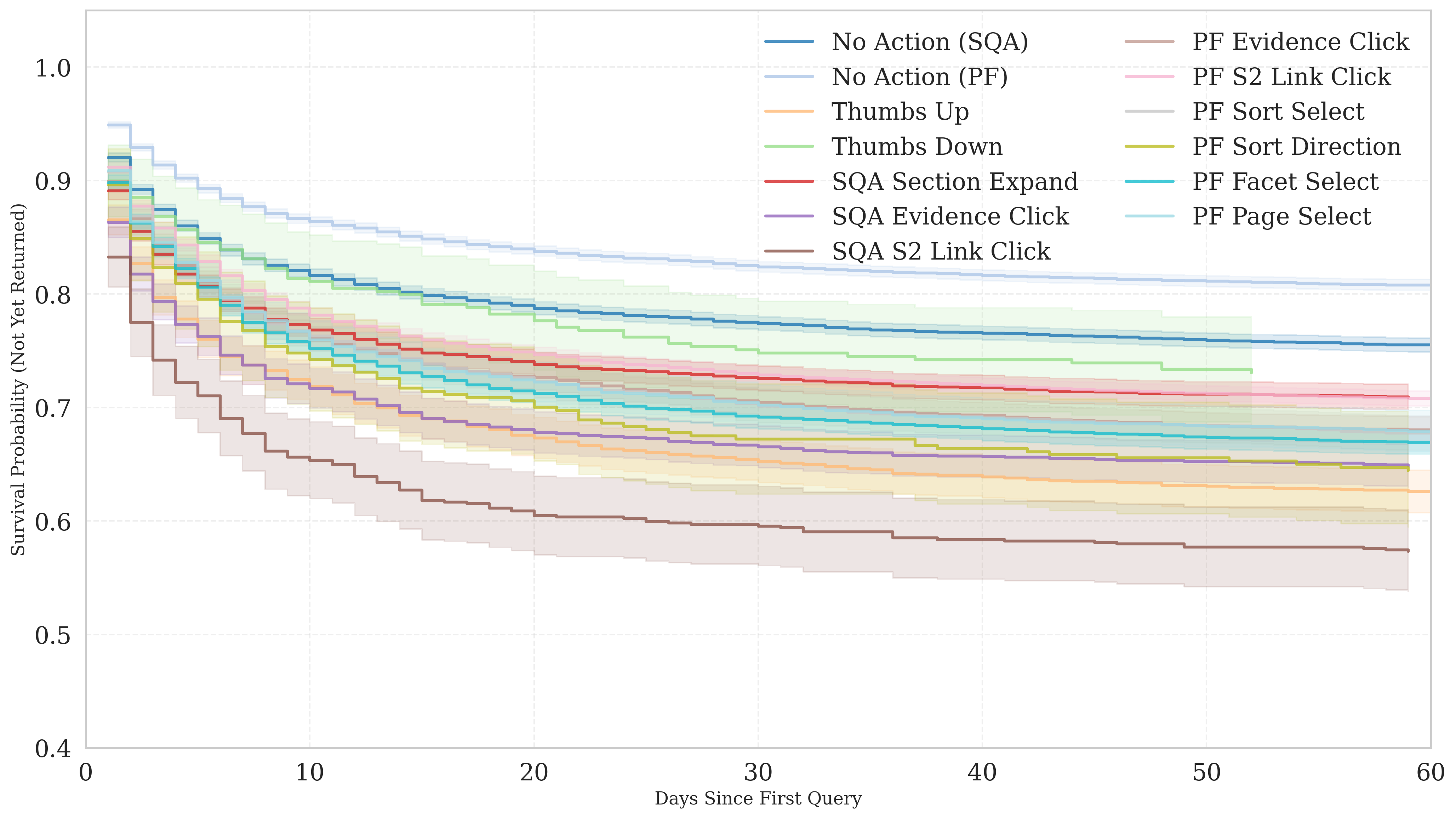}
  \caption{Survival analysis showing the Kaplan-Meier curve associated
    with various actions that users performed during their first visit
    to \sys. Users who did not perform any action take longer to
    return to \sys than those who do.}
  \label{fig:action-survival}
\end{figure}

\begin{figure*}[htbp]
  \centering
  \begin{subfigure}[t]{0.48\linewidth}
    \centering
    \includegraphics[width=\linewidth]{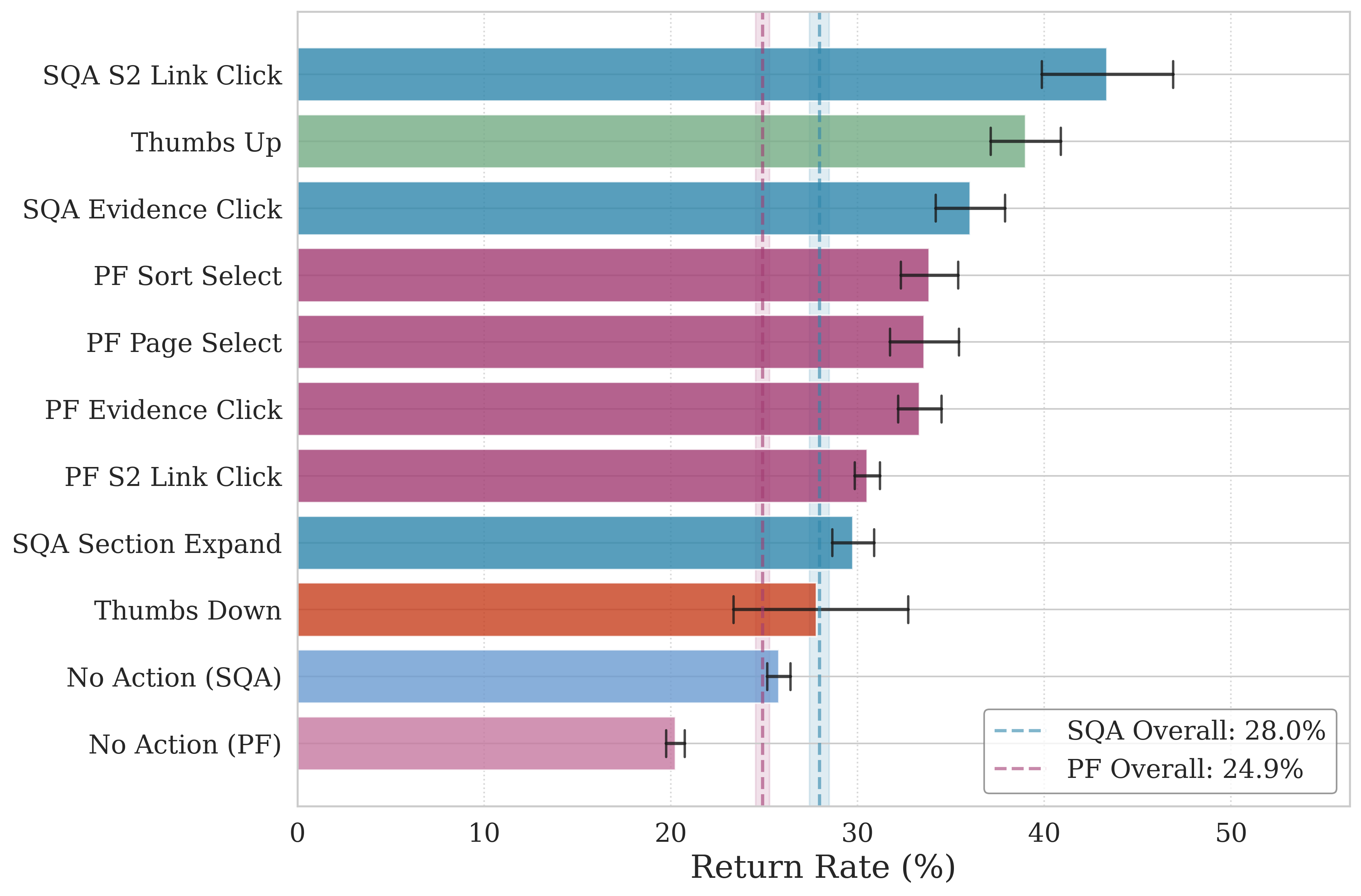}
    \caption{Return rates of first time users. Users who perform
      thumbs up are likely to return, users who don't perform any
      action are the least likely to return.}
    \label{fig:action-return-first}
  \end{subfigure}
  \hfill
  \begin{subfigure}[t]{0.48\linewidth}
    \centering
    \includegraphics[width=\linewidth]{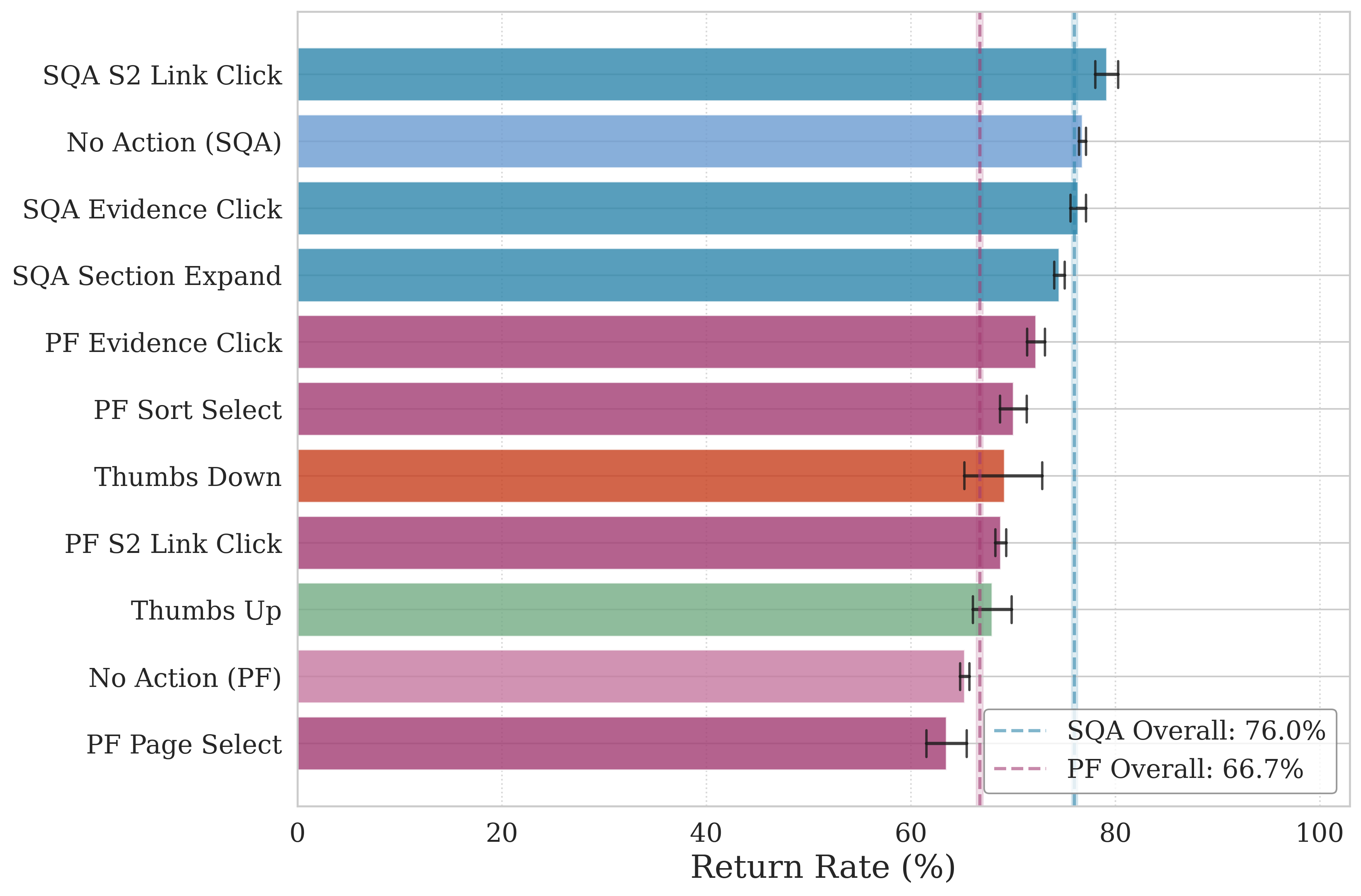}
    \caption{Return rates for users after they have initiated at least
      one query. Link clicks are at least as good an indicator of
      return as thumbs up. \texttt{Page select}, an action which
      suggests frustration, is associated with non-return.}
    \label{fig:action-return-thread2plus}
  \end{subfigure}
  \caption{Action return rates showing the probability of users
    returning after different action types. Actions of first time
    users are likely more affected by novelty effect.}
  \label{fig:action-return-rates}
\end{figure*}

\begin{table}[t]
  \centering
  \footnotesize
  \caption{Example \pf queries by type. \pf performs
    well on Citation \& Evidence Finding and Broad Topic Exploration
    queries but struggles with queries having Temporal Constraints,
    Data/Resource requirements, or Content Generation requests.}
  \label{tab:pf-query-examples}
  \begin{tabularx}{\linewidth}{lX}
    \toprule
    \textbf{Query Type} & \textbf{Example Query} \\
    \midrule
    Citation \& Evidence Finding & \textit{Can you find for me explicitly mentioned in a paper when an anti-reflection coating was used to improve the temporal contrast of a high power laser due to it having to be at s polarization.} \\
    \addlinespace
    Broad Topic Exploration & \textit{find me articles on AI implementation in SMEs.} \\
    \addlinespace
    Temporal Constraints & \textit{What are the drawbacks of Intelligent tutoring systems from before they used LLMs} \\
    \addlinespace
    Data/Resource Availability & \textit{dataset with image and sun azimuth and sun elevation} \\
    \addlinespace
    Content Generation and Expansion & \textit{Create an article with the title ``Material Strength Design on Excavator Arm Model'' with references of 40 articles or journals} \\
    \bottomrule
  \end{tabularx}
\end{table}

\begin{table}[t]
  \centering
  \footnotesize
  \caption{Example \sqa queries by type. \sqa excels at
    Concept Definition and Complex Contextual Narrative queries but
    struggles with Citation Format Specification requests.}
  \label{tab:sqa-query-examples}
  \begin{tabularx}{\linewidth}{lX}
    \toprule
    \textbf{Query Type} & \textbf{Example Query} \\
    \midrule
    Concept Definition and Explanation & \textit{explain what
                                         multimodal and multisensor
                                         are and the differents betwen
                                         them} \\
    \addlinespace
    Complex Contextual Narrative & \textit{Recent evidence highlights
                                   the critical role of growth hormone
                                   secretagogue receptor signaling in
                                   hippocampal synaptic
                                     physiology, mediated through
                                   dopamine receptor
                                   activity... [describes detailed
                                   molecular mechanisms]} \\
    \addlinespace
    Citation Format Specification & \textit{Use APA style 7 for
                                    in-text citations, citing authors'
                                    names, publication dates, and page
                                    numbers. Write a theoretical
                                      framework of 17,000 words about
                                    schooling of rural girls...} \\
    \bottomrule
  \end{tabularx}
\end{table}

\subsection{Query characteristics and success metrics}
\label{app:query-success}

To understand which types of queries current systems handle well and where they struggle, we fit logistic regression models predicting click-through rate (CTR) from query intent, phrasing style, search criteria, and field of study, with user features as controls (see \Cref{app:ctr-validation} for CTR validation). \Cref{tab:pf-ctr-coef-appendix,tab:sqa-ctr-coef-appendix} report the resulting odds ratios (OR $>$ 1 indicates higher click odds). We focus on query aspects that are independent of field of study (intent, phrasing, and criteria) because these have direct implications for system design, whereas field effects likely reflect properties of the user population or corpus coverage (see \Cref{tab:pf-field-coef} for statistically significant field coefficients).

Since \pf and \sqa are designed for different purposes, we analyze each tool separately (see \Cref{tab:pf-query-examples,tab:sqa-query-examples} for example queries).

\pf{} struggles with \texttt{Content Generation and Expansion} queries (which it is not designed for) and queries with \texttt{Temporal Constraints} or \texttt{Data Resource} requirements such as publicly available datasets (which may be unsatisfiable). On the flip side, \pf{} performs well on queries it was designed for: \texttt{Citation/Evidence Finding} and \texttt{Broad Topic Exploration} queries both have higher click odds.

\sqa{} shows an analogous pattern. \texttt{Concept Definition and Explanation} queries have higher click odds, as \sqa{} was designed with them in mind. \texttt{Complex Contextual Narrative} queries also have high click odds, suggesting users achieve success with complex queries that would typically fail using traditional IR tools. \texttt{Citation Format Specification} queries have lower click odds because \sqa{} uses a single fixed citation format rather than adapting to user-specified styles.

\begin{table*}[htbp]
  \centering
  \footnotesize
  \begin{minipage}[t]{0.48\linewidth}
    \centering
    \caption{Odds ratios for \sTwo link click on \pf
      (95\% CI; user features included as controls but
      not shown). \texttt{Content Gen.\ \& Exp.} queries have
      approximately half the odds of a click (OR\,=\,0.47), while
      \texttt{Cit./Evid.\ Finding} queries have 17\% higher odds
      (OR\,=\,1.17). Field of study coefficients are shown in
      \Cref{tab:pf-field-coef}. Only statistically significant effects
    are presented.}
    \label{tab:pf-ctr-coef-appendix}
    \begin{tabular}{@{}lcr@{}}
      \toprule
      \textbf{Predictor} & \textbf{Group} & \textbf{OR} \\
      \midrule
      Cit./Evid.\ Finding & I & \ci{1.17}{0.13}{0.15} \\
      Broad Topic Expl. & I & \ci{1.12}{0.08}{0.08} \\
      Temporal Const. & C & \ci{0.82}{0.11}{0.12} \\
      Data Res.\ Avail. & C & \ci{0.61}{0.14}{0.18} \\
      Content Gen.\ \& Exp. & I & \ci{0.47}{0.16}{0.24} \\
      \bottomrule
    \end{tabular}
    \\[0.5ex]
    {\scriptsize Group: I=intent, C=criteria. OR=Odds Ratio.}
  \end{minipage}
  \hfill
  \begin{minipage}[t]{0.48\linewidth}
    \centering
    \caption{Odds ratios for \sTwo link click on \sqa
      (95\% CI; user features included as controls but
      not shown). Fewer predictors reach significance compared to \pf.
      \texttt{Complex Ctx.\ Narr.} queries have 47\% higher
      odds of a click (OR\,=\,1.47), while \texttt{Cit.\ Format
        Spec.} queries have 38\% lower odds (OR\,=\,0.62).}
    \label{tab:sqa-ctr-coef-appendix}
    \begin{tabular}{@{}lcr@{}}
      \toprule
      \textbf{Predictor} & \textbf{Group} & \textbf{OR} \\
      \midrule
      Complex Ctx.\ Narr. & P & \ci{1.47}{0.31}{0.39} \\
      Concept Def.\ \& Expl. & I & \ci{1.29}{0.15}{0.17} \\
      Cit.\ Format Spec. & P & \ci{0.62}{0.18}{0.25} \\
      \bottomrule
    \end{tabular}
    \\[0.5ex]
    {\scriptsize Group: I=intent, P=phrasing. OR=Odds Ratio.}
  \end{minipage}
\end{table*}

\begin{figure*}[htbp]
  \centering
  \begin{subfigure}[t]{0.48\linewidth}
    \centering
    \includegraphics[width=\linewidth]{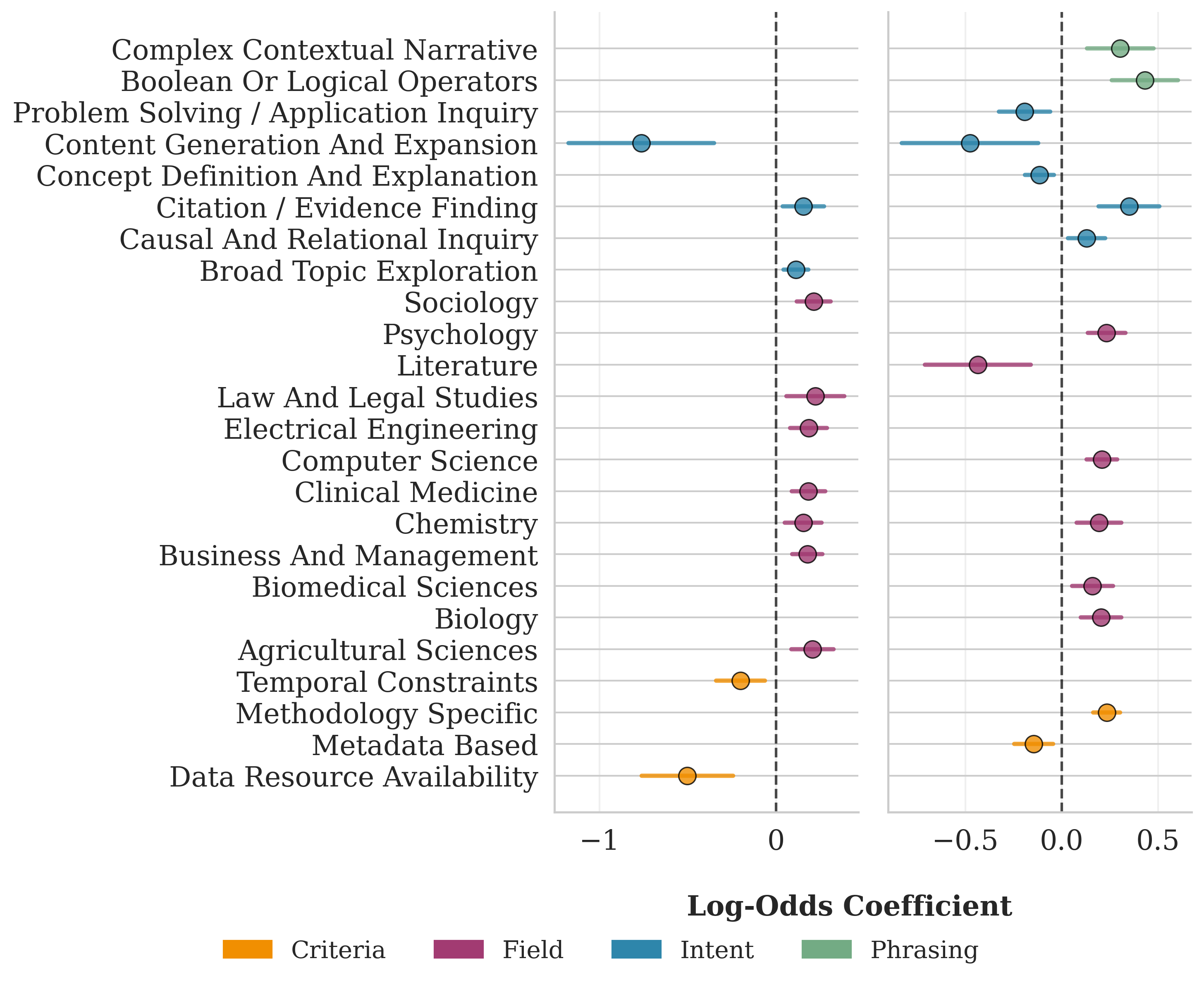}
    \caption{Coefficients for \pf model predicting S2 clicks (left) and
      user return (right).}
    \label{fig:pf-coef-appendix}
  \end{subfigure}
  \hfill
  \begin{subfigure}[t]{0.48\linewidth}
    \centering
    \includegraphics[width=\linewidth]{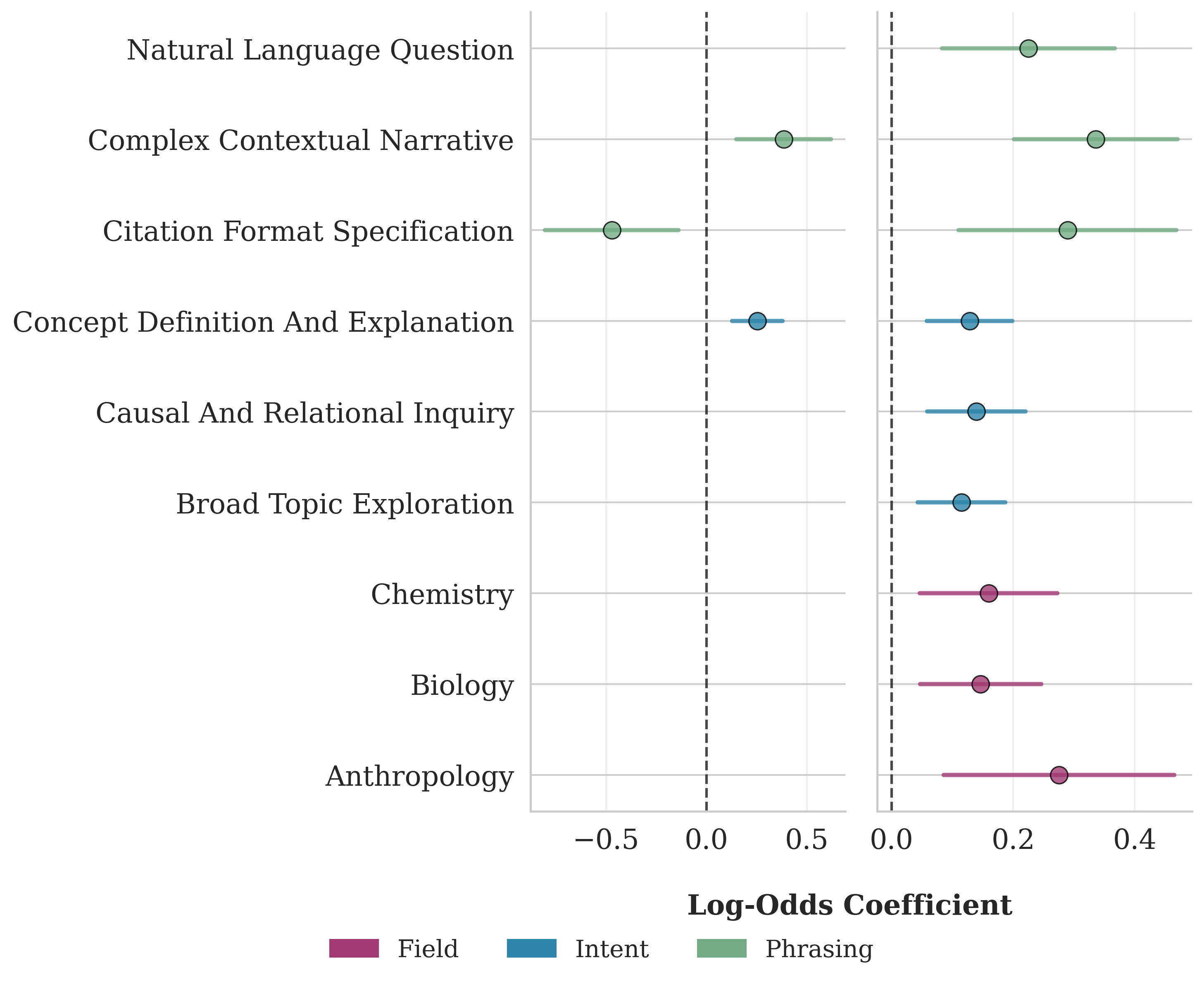}
    \caption{Coefficients for \sqa model predicting S2 clicks (left) and
      user return (right).}
    \label{fig:sqa-coef-appendix}
  \end{subfigure}
  \caption{Estimated coefficients for the linear models predicting
    clicks and return on \pf and \sqa. Coefficients are shown with 95\%
    confidence intervals. Only coefficients significant after
    Benjamini-Hochberg correction are displayed.}
  \label{fig:success-metrics-comparison-appendix}
\end{figure*}

\begin{table}[htbp]
  \centering
  \small
  \caption{Field of study odds ratios for \sTwo link click on \pf (95\% CI).}
  \label{tab:pf-field-coef}
  \begin{tabular}{@{}lr@{}}
    \toprule
    \textbf{Field of Study} & \textbf{OR} \\
    \midrule
    Law \& Legal Studies & \ci{1.25}{0.19}{0.22} \\
    Sociology & \ci{1.24}{0.11}{0.12} \\
    Agricultural Sciences & \ci{1.23}{0.14}{0.16} \\
    Electrical Engineering & \ci{1.20}{0.12}{0.13} \\
    Clinical Medicine & \ci{1.20}{0.11}{0.12} \\
    Business \& Management & \ci{1.20}{0.10}{0.11} \\
    Chemistry & \ci{1.17}{0.12}{0.13} \\
    \bottomrule
  \end{tabular}
\end{table}

\begin{table}[htbp]
  \centering
  \small
  \caption{User return rate by LLM quality assessment of \sqa reports.
    Users who received reports assessed as high-quality by an LLM are
    significantly more likely to return ($p < 0.001$, $r = 0.052$).}
  \label{tab:return-vs-llm-quality}
  \begin{tabular}{@{}lrrr@{}}
    \toprule
    \textbf{LLM Assessment} & \textbf{$n$} & \textbf{Return Rate} & \textbf{95\% CI} \\
    \midrule
    Negative & 1,275 & 54.7\% & [52.0\%, 57.5\%] \\
    Positive & 8,725 & 62.3\% & [61.3\%, 63.3\%] \\
    \bottomrule
  \end{tabular}
\end{table}

\begin{table}[htbp]
  \centering
  \small
  \caption{\sTwo link click rate by LLM quality assessment of \sqa reports.
    Users who received reports assessed as high-quality by an LLM are
    more likely to click on citation links ($p = 0.01$, $r = 0.026$).}
  \label{tab:s2-clicks-vs-llm-quality}
  \begin{tabular}{@{}lrrr@{}}
    \toprule
    \textbf{LLM Assessment} & \textbf{$n$} & \textbf{Click Rate} & \textbf{95\% CI} \\
    \midrule
    Negative & 1,373 & 3.9\% & [3.0\%, 5.0\%] \\
    Positive & 8,631 & 5.5\% & [5.1\%, 6.0\%] \\
    \bottomrule
  \end{tabular}
\end{table}

\begin{table}[htbp]
  \centering
  \small
  \caption{Confusion matrix comparing user feedback (thumbs up/down)
    with LLM quality assessment. The LLM achieves 73.9\% accuracy,
    with higher agreement on positive assessments (F1 = 0.83) than
    negative ones (F1 = 0.47), reflecting the challenge of identifying
    subtle quality issues.}
  \label{tab:feedback-vs-llm-confusion}
  \begin{tabular}{@{}lrr@{}}
    \toprule
    & \multicolumn{2}{c}{\textbf{LLM Assessment}} \\
    \cmidrule(l){2-3}
    \textbf{User Feedback} & Positive & Negative \\
    \midrule
    Thumbs Up & 86 & 14 \\
    Thumbs Down & 22 & 16 \\
    \bottomrule
  \end{tabular}
\end{table}


%% file: appendix_dataset.tex
\section{Dataset Schema}
\label{app:schema}

The released dataset comprises six parquet files. Each query generates a report, identified by its \texttt{thread\_id}. All files can be joined on \texttt{thread\_id}.

\subsection{optin\_queries\_anonymized.parquet}
User queries submitted to \sqa and \pf.

\begin{tabular}{lll}
\toprule
\textbf{Column} & \textbf{Type} & \textbf{Description} \\
\midrule
\texttt{query} & string & The user's query text \\
\texttt{thread\_id} & string & Hashed report identifier \\
\texttt{query\_ts} & timestamp & Query submission time \\
\texttt{tool} & string & Tool used: \texttt{sqa} or \texttt{pf} \\
\bottomrule
\end{tabular}

\subsection{section\_expansions\_anonymized.parquet}
Records of users expanding sections in \sqa reports (sections are collapsed by default).

\begin{tabular}{lll}
\toprule
\textbf{Column} & \textbf{Type} & \textbf{Description} \\
\midrule
\texttt{thread\_id} & string & Hashed report identifier \\
\texttt{section\_expand\_ts} & timestamp & Expansion time \\
\texttt{section\_id} & int & Index of expanded section (0-indexed) \\
\bottomrule
\end{tabular}

\subsection{s2\_link\_clicks\_anonymized.parquet}
Clicks on Semantic Scholar paper links within either tool.

\begin{tabular}{lll}
\toprule
\textbf{Column} & \textbf{Type} & \textbf{Description} \\
\midrule
\texttt{thread\_id} & string & Hashed report identifier \\
\texttt{s2\_link\_click\_ts} & timestamp & Click time \\
\texttt{corpus\_id} & int & Semantic Scholar corpus ID \\
\texttt{tool} & string & Tool: \texttt{sqa} or \texttt{pf} \\
\bottomrule
\end{tabular}

\subsection{report\_section\_titles\_anonymized.parquet}
Section titles from \sqa generated reports.

\begin{tabular}{lll}
\toprule
\textbf{Column} & \textbf{Type} & \textbf{Description} \\
\midrule
\texttt{thread\_id} & string & Hashed report identifier \\
\texttt{section\_idx} & int & Section index (0-indexed) \\
\texttt{section\_title} & string & Title of the section \\
\bottomrule
\end{tabular}

\subsection{report\_corpus\_ids\_anonymized.parquet}
Papers cited in \sqa report sections.

\begin{tabular}{lll}
\toprule
\textbf{Column} & \textbf{Type} & \textbf{Description} \\
\midrule
\texttt{thread\_id} & string & Hashed report identifier \\
\texttt{corpus\_id} & int & Semantic Scholar corpus ID \\
\bottomrule
\end{tabular}

\subsection{pf\_shown\_results\_anonymized.parquet}
Papers shown in \pf search results.

\begin{tabular}{lll}
\toprule
\textbf{Column} & \textbf{Type} & \textbf{Description} \\
\midrule
\texttt{thread\_id} & string & Hashed report identifier \\
\texttt{query\_ts} & timestamp & Query submission time \\
\texttt{result\_position} & int & Position in results list (0-indexed) \\
\texttt{corpus\_id} & int & Semantic Scholar corpus ID \\
\bottomrule
\end{tabular}

%% file: appendix_taxonomies.tex
\section{Label definitions and few-shot examples}
\label{app:labels}

This section provides comprehensive definitions for all query intents, search criteria, and phrasing styles identified in our taxonomy for both ScholarQA (SQA) and Paper Finder (PF). Each category includes a description and representative examples from actual user queries. Query intent taxonomies are presented in Tables~\ref{tab:sqa-intents-full}--\ref{tab:sqa-intents-full-2} (SQA) and Tables~\ref{tab:pf-intents-full}--\ref{tab:pf-intents-full-2} (PF). Search criteria taxonomies are presented in Table~\ref{tab:sqa-criteria-full} (SQA) and Table~\ref{tab:pf-criteria-full} (PF). Query phrasing taxonomies are presented in Table~\ref{tab:sqa-phrasing-full} (SQA) and Table~\ref{tab:pf-phrasing-full} (PF). The 28 fields of study used for classification are listed in Table~\ref{tab:fields}.

\begin{table*}[t]
\centering
\footnotesize
\begin{tabular}{p{0.22\textwidth}p{0.38\textwidth}p{0.35\textwidth}}
\toprule
\textbf{Intent} & \textbf{Description} & \textbf{Examples} \\
\midrule
Broad Topic Exploration & The user wants a general overview, literature review, or summary of a broad academic field or topic without highly specific constraints. & \textit{``Conduct a literature review on the use of deep learning in focussed ultrasound stimulation''} \newline \textit{``E-commerce''} \newline \textit{``Theories of internalization''} \\
\addlinespace
Specific Factual Retrieval & The user is asking for a precise, factual piece of information, a specific data point, a statistic, or a historical detail. This intent is characterized by its specificity and aims for a direct, verifiable answer. & \textit{``what is the best wavelength to check the hemolysis in plasma samples?''} \newline \textit{``What was the first study examine sexual double standards? When was the term coined?''} \newline \textit{``find user statistics for UniProt''} \\
\addlinespace
Concept Definition and Explanation & The user seeks to understand the meaning, core principles, and fundamental aspects of a specific academic theory, concept, model, or term. & \textit{``What are reasoning LLMs?''} \newline \textit{``overview of the Scholarly Primitive Theory by John unsworth''} \newline \textit{``What is `Administration' in Public Financial Administration''} \\
\addlinespace
Comparative Analysis & The user wants to understand the similarities, differences, advantages, and disadvantages between two or more concepts, theories, models, or methods. & \textit{``employer brand vs employer branding''} \newline \textit{``What are the differences and similarity between information foraging theory and exploratory search?''} \newline \textit{``Compare the concept of world order between the theory of realism, liberalism, constructivism...''} \\
\addlinespace
Methodological and Procedural Guidance & The user is looking for instructions, protocols, frameworks, or best practices for carrying out a specific research task, experiment, analysis, or procedure. & \textit{``what can I do to parse effectively PDFs for RAG?''} \newline \textit{``how to extract dna from blood for sepsis molecular diagnosis''} \newline \textit{``What are the best research questions related to the following research Gap?...''} \\
\bottomrule
\end{tabular}
\caption{Complete ScholarQA query intent taxonomy (part 1 of 3).}
\label{tab:sqa-intents-full}
\end{table*}

\begin{table*}[t]
\centering
\footnotesize
\begin{tabular}{p{0.22\textwidth}p{0.38\textwidth}p{0.35\textwidth}}
\toprule
\textbf{Intent} & \textbf{Description} & \textbf{Examples} \\
\midrule
Tool and Resource Discovery & The user is searching for specific academic resources such as datasets, software tools, questionnaires, evaluation benchmarks, or code repositories relevant to their research. & \textit{``Question answering datasets''} \newline \textit{``find the deepfake dataset which contains some real-world perturbations.''} \newline \textit{``What is the best questionnaire to measure mathematics anxiety?''} \\
\addlinespace
Research Gap and Limitation Analysis & The user aims to identify the limitations, unsolved problems, or unexplored areas (gaps) within a specific field of research, often to justify a new study or find a novel research direction. & \textit{``Systematic reviews on AWE tools in EFL argumentative writing. What is the gap?''} \newline \textit{``Limitations: What limitations are present in the theoretical literature on spatial poverty?''} \newline \textit{``Review the SOTA and research gaps of organic neuromorphic computing in AI...''} \\
\addlinespace
Causal and Relational Inquiry & The user wants to understand the relationship, impact, effect, or influence of one variable or concept on another. These queries often explore cause-and-effect connections or correlations. & \textit{``Does sleep consistency benefit overall health''} \newline \textit{``how do CEO power influence sustainability performance and financial performance of companies?''} \newline \textit{``drivers of public support for cost-intensive policies''} \\
\addlinespace
Focused Academic Synthesis & The user requests a structured and constrained synthesis of academic literature, such as a systematic review, a meta-analysis, or a review limited to specific journals, timeframes, or methodologies. & \textit{``Future of the Labor Market and Higher Education `Only use articles from Q1 and Q2 journals.'''} \newline \textit{``Write a systematic review on soybean production challenges and opportunities in Ethiopia''} \newline \textit{``Generate a comprehensive, citation-rich literature review (2014–2025)...''} \\
\addlinespace
Academic Document Drafting & The user explicitly asks for the generation of a specific academic document or a section of one, such as a proposal, a chapter, an abstract, or a detailed review, often providing a structured template or detailed instructions. & \textit{``using APA 7 style with in text citation, write a literature review of about 2000 words...''} \newline \textit{``make me a study justification for Impacted canines and its association with dental anomalies...''} \newline \textit{``write a comprehensive theoretical framework for attention and sustained attention''} \\
\bottomrule
\end{tabular}
\caption{Complete ScholarQA query intent taxonomy (part 2 of 3).}
\label{tab:sqa-intents-full-1b}
\end{table*}

\begin{table*}[t]
\centering
\footnotesize
\begin{tabular}{p{0.22\textwidth}p{0.38\textwidth}p{0.35\textwidth}}
\toprule
\textbf{Intent} & \textbf{Description} & \textbf{Examples} \\
\midrule
Ideation & The user presents a complex problem, a scenario, or a research objective and asks for potential solutions, innovative ideas, hypotheses, or new research directions. & \textit{``I want to do `quantization in MoE model inference', find some related papers and recommend some ideas on this''} \newline \textit{``how can the anthropomorphization of artificial intelligence influence teaching practices...''} \newline \textit{``Has any research been done to more fully articulate the collaboration/cooperation required...''} \\
\addlinespace
Application Inquiry & User seeks research focused on solving a specific, practical problem or demonstrating the application of a theory or technology in a real-world context. & \textit{``The Effect of a Program Based on Using Zoom Application on Developing Level Two Students' Business Vocabulary...''} \newline \textit{``Golf shot/swing analysis and feedback for self training''} \newline \textit{``research paper on benefit of construction machineries application''} \\
\addlinespace
Data Interpretation and Analysis Support & The user needs help with interpreting or analyzing specific data, results, or observations. This can include understanding statistical outputs, analyzing images/spectra, or making sense of experimental findings. & \textit{``analyze UV-Visible peak of HDPE at 281 nm''} \newline \textit{``What is the clinical and biological implication and meaning of the blood and nail Selenium levels...''} \newline \textit{``analyzed uv visible spectrum of Fe3O4 nanopartticles at 262 nm''} \\
\addlinespace
Content Generation and Expansion & User provides a piece of existing text (such as an outline, abstract, or draft) and explicitly requests for it to be expanded, rewritten, or to have components like references generated for it. & \textit{``My research topic is `The Effects of Virtual Reality Environments on Adult English Language Acquisition.' Personalize the following questions...''} \newline \textit{``melhore essa analise A Figura exibe dois picos expressivos...''} \newline \textit{``The method of chemical coprecipitation allows to obtain barium hexaferrite... select literary sources...''} \\
\addlinespace
Complex Cross-Paper Synthesis & The user requires complex reasoning that involves synthesizing information across multiple research papers to understand relationships, trace evolution of concepts, identify emergent patterns, or connect insights from different disciplines. & \textit{``How has the conditional Pesin entropy formula evolved in the literature on coupled dynamical systems...''} \newline \textit{``Has any research been done to more fully articulate the collaboration/cooperation required...''} \newline \textit{``topological structure of knowledge networks or knowledge maps...''} \\
\addlinespace
Citation \& Evidence Finding & User provides a specific claim, statement, or fact and requests academic sources that can be used as a citation to support it. & \textit{``Please search for literature to prove that BAK is mainly distributed on the outer membrane of mitochondria''} \newline \textit{``it was shown that hydroxyl groups... can form intramolecular hydrogen bonds... find me a citation for that''} \newline \textit{``Media video dan televisi adalah media pembelajaran... carikan sumbernya''} \\
\addlinespace
Specific Paper Retrieval & User attempts to locate a single, known academic paper, often using a DOI, a partial citation, or specific author and title information. & \textit{``10.1097/PRS.0000000000008882''} \newline \textit{``To improve the adaptability of control systems, researchers have investigated... Amiri et al.... find this paper''} \newline \textit{``Smith J, et al. The regulation of sperm motility... [citation]''} \\
\bottomrule
\end{tabular}
\caption{Complete ScholarQA query intent taxonomy (part 3 of 3). Taxonomy was built through iterative human review and LLM-assisted analysis using Gemini-2.5-pro on a sample of 1000 queries.}
\label{tab:sqa-intents-full-2}
\end{table*}

\begin{table*}[t]
\centering
\footnotesize
\begin{tabular}{p{0.22\textwidth}p{0.38\textwidth}p{0.35\textwidth}}
\toprule
\textbf{Intent} & \textbf{Description} & \textbf{Examples} \\
\midrule
Broad Topic Exploration & The user wants a general overview, literature review, or summary of a broad academic field or topic without highly specific constraints. & \textit{``Do a literature review on artificial intelligence in management accounting''} \newline \textit{``papers about the technological advances, debates and challenges in humanitarian aid''} \newline \textit{``Show some exploratory qualitative studies on ELT''} \\
\addlinespace
Specific Paper Retrieval & User attempts to locate a single, known academic paper, often using a DOI, a partial citation, or specific author and title information. & \textit{``10.1097/PRS.0000000000008882''} \newline \textit{``To improve the adaptability of control systems... Amiri et al.... find this paper''} \newline \textit{``Smith J, et al. The regulation of sperm motility... [citation]''} \\
\addlinespace
Complex Cross-Paper Synthesis & The user requires complex reasoning that involves synthesizing information across multiple research papers to understand relationships, trace evolution of concepts, identify emergent patterns, or connect insights from different disciplines. & \textit{``How has the conditional Pesin entropy formula evolved in the literature on coupled dynamical systems...''} \newline \textit{``Has any research been done to more fully articulate the collaboration/cooperation required...''} \newline \textit{``topological structure of knowledge networks...''} \\
\addlinespace
Citation \& Evidence Finding & User provides a specific claim, statement, or fact and requests academic sources that can be used as a citation to support it. & \textit{``Please search for literature to prove that BAK is mainly distributed on the outer membrane of mitochondria''} \newline \textit{``it was shown that hydroxyl groups... find me a citation for that''} \newline \textit{``Media video dan televisi... carikan sumbernya''} \\
\addlinespace
Methodological and Procedural Guidance & The user is looking for instructions, protocols, frameworks, or best practices for carrying out a specific research task, experiment, analysis, or procedure. & \textit{``How to screen the downstream pathways of ERK''} \newline \textit{``How is alignment consistency ensured from part identification to measurement feedback...''} \newline \textit{``what is the importance of covering many different spatial configurations...''} \\
\bottomrule
\end{tabular}
\caption{Complete Paper Finder query intent taxonomy (part 1 of 3).}
\label{tab:pf-intents-full}
\end{table*}

\begin{table*}[t]
\centering
\footnotesize
\begin{tabular}{p{0.22\textwidth}p{0.38\textwidth}p{0.35\textwidth}}
\toprule
\textbf{Intent} & \textbf{Description} & \textbf{Examples} \\
\midrule
Concept Definition and Explanation & The user seeks to understand the meaning, core principles, and fundamental aspects of a specific academic theory, concept, model, or term. & \textit{``ppf''} \newline \textit{``Definition of climate''} \newline \textit{``What is artificial intelligence?''} \\
\addlinespace
Comparative Analysis & The user wants to understand the similarities, differences, advantages, and disadvantages between two or more concepts, theories, models, or methods. & \textit{``find paper that comparing the American native or indigenous or village chicken egg and commercial chicken egg''} \newline \textit{``Copare the papers of learning analytics through UTAUT and TAM''} \newline \textit{``Comparison of the efficiency of in-vitro digestibility to in-vivo digestibility''} \\
\addlinespace
Tool and Resource Discovery & The user is searching for specific academic resources such as datasets, software tools, questionnaires, evaluation benchmarks, or code repositories relevant to their research. & \textit{``Papers introducing a dataset of an unscripted dialogue between 2 speakers...''} \newline \textit{``i need Research papers and open datasets on sports injuries... sensor data such as fitbit...''} \newline \textit{``Document Processing LLM Benchmarks''} \\
\addlinespace
Causal and Relational Inquiry & The user wants to understand the relationship, impact, effect, or influence of one variable or concept on another. These queries often explore cause-and-effect connections or correlations. & \textit{``How does fear of missing out influence the effect of mindfulness on academic procrastination...''} \newline \textit{``The Impact of Corporate Governance Practices on Financial Inclusion...''} \newline \textit{``what are sex differences in psychiatric comorbidities among individuals with migraine?''} \\
\addlinespace
Application Inquiry & User seeks research focused on solving a specific, practical problem or demonstrating the application of a theory or technology in a real-world context. & \textit{``The Effect of a Program Based on Using Zoom Application on Developing Level Two Students' Business Vocabulary...''} \newline \textit{``Golf shot/swing analysis and feedback for self training''} \newline \textit{``research paper on benefit of construction machineries application''} \\
\bottomrule
\end{tabular}
\caption{Complete Paper Finder query intent taxonomy (part 2 of 3).}
\label{tab:pf-intents-full-1b}
\end{table*}

\begin{table*}[t]
\centering
\footnotesize
\begin{tabular}{p{0.22\textwidth}p{0.38\textwidth}p{0.35\textwidth}}
\toprule
\textbf{Intent} & \textbf{Description} & \textbf{Examples} \\
\midrule
Focused Academic Synthesis & The user requests a structured and constrained synthesis of academic literature, such as a systematic review, a meta-analysis, or a review limited to specific journals, timeframes, or methodologies. & \textit{``emerging trends in communications sectors that Ofcom deals with''} \newline \textit{``recent advances in nuclear fusion power''} \newline \textit{``Must read papers for LLM''} \\
\addlinespace
Content Generation and Expansion & User provides a piece of existing text (such as an outline, abstract, or draft) and explicitly requests for it to be expanded, rewritten, or to have components like references generated for it. & \textit{``My research topic is `The Effects of Virtual Reality Environments on Adult English Language Acquisition.' Personalize the following questions...''} \newline \textit{``melhore essa analise...''} \newline \textit{``The method of chemical coprecipitation... select literary sources...''} \\
\addlinespace
Data Interpretation and Analysis Support & The user needs help with interpreting or analyzing specific data, results, or observations. This can include understanding statistical outputs, analyzing images/spectra, or making sense of experimental findings. & \textit{``analyze UV-Visible peak of HDPE at 281 nm''} \newline \textit{``What is the clinical and biological implication and meaning of the blood and nail Selenium levels...''} \newline \textit{``analyzed uv visible spectrum of Fe3O4 nanopartticles at 262 nm''} \\
\addlinespace
Ideation & The user presents a complex problem, a scenario, or a research objective and asks for potential solutions, innovative ideas, hypotheses, or new research directions. & \textit{``I want to do `quantization in MoE model inference', find some related papers and recommend some ideas on this''} \newline \textit{``how can the anthropomorphization of artificial intelligence influence teaching practices...''} \newline \textit{``Has any research been done to more fully articulate the collaboration/cooperation required...''} \\
\addlinespace
Research Gap and Limitation Analysis & The user aims to identify the limitations, unsolved problems, or unexplored areas (gaps) within a specific field of research, often to justify a new study or find a novel research direction. & \textit{``Systematic reviews on AWE tools in EFL argumentative writing. What is the gap?''} \newline \textit{``Limitations: What limitations are present in the theoretical literature on spatial poverty?''} \newline \textit{``Review the SOTA and research gaps of organic neuromorphic computing in AI...''} \\
\addlinespace
Specific Factual Retrieval & The user is asking for a precise, factual piece of information, a specific data point, a statistic, or a historical detail. This intent is characterized by its specificity and aims for a direct, verifiable answer. & \textit{``what is the best wavelength to check the hemolysis in plasma samples?''} \newline \textit{``What was the first study examine sexual double standards? When was the term coined?''} \newline \textit{``find user statistics for UniProt''} \\
\bottomrule
\end{tabular}
\caption{Complete Paper Finder query intent taxonomy (part 3 of 3). Taxonomy was built through iterative human review and LLM-assisted analysis using Gemini-2.5-pro on a sample of 1000 queries.}
\label{tab:pf-intents-full-2}
\end{table*}

\begin{table*}[t]
\centering
\footnotesize
\begin{tabular}{p{0.22\textwidth}p{0.38\textwidth}p{0.35\textwidth}}
\toprule
\textbf{Category} & \textbf{Description} & \textbf{Examples} \\
\midrule
Methodology-Specific Criteria & Queries where the user requires specific methods, approaches, analytical techniques, or experimental designs to be present in the search results. This can include demands for computational models, experimental paradigms, or meta-analyses. & \textit{``Fe dopped ZnO and its composite with zeolite, explain crystallite size by Scherrer and Williamson-Hall models and compare''} \newline \textit{``meta analysis on behavioral epigenetics dogs and cats...''} \newline \textit{``compare las membranas poliméricas de inclusión respecto a cada uno de los métodos convencionales...''} \\
\addlinespace
Publication Type/Quality Filters & Queries that require certain types of literature based on format, peer review status, or journal ranking, such as papers, theses, reviews, or guidelines, with or without requirements for high-impact or reputable sources. & \textit{``Review papers on SOTA and research gaps in manifold learning in AI.''} \newline \textit{``provide several examples from original research articles published post-2020, excluding review articles''} \newline \textit{``find articles suggesting a research gap on the influence of OFW parentage...''} \\
\addlinespace
Temporal Constraints & Queries that include requirements related to the publication date or period of the information (e.g., recent literature, post-2020, historical periods). This includes searching within specific timeframes or for recent advances. & \textit{``Find policies and events in European countries that demonstrate `strategic autonomy' between 2016 and 2025.''} \newline \textit{``provide several examples from original research articles published post-2020...''} \newline \textit{``Latest research on social loneliness and perfectionism''} \\
\addlinespace
Metadata-Based Criteria & Queries specifying attributes of the publications or their creators, such as language, authorship, institutional affiliation, location/country, or publication venue (journal, conference, etc.). & \textit{``in welchen Studien wurde das R paket treeclim verwendet und was wurde damit gemacht?''} \newline \textit{``Shahinda Rezk''} \newline \textit{``search for papers published in NeurIPS about reinforcement learning''} \\
\addlinespace
Citation and Impact-Based Criteria & Queries that specify requirements related to citation counts, impact factors, highly-cited papers, or seminal works in a field. This can include searching for influential or foundational studies. & \textit{``find the most highly cited papers on deep learning optimization''} \newline \textit{``seminal works on contract-net protocols''} \newline \textit{``SOTA metric on the canonical benchmark(s) and the delta or change vs. prior best benchmark score''} \\
\addlinespace
Data/Resource Availability Criteria & Queries that ask for studies or papers with available datasets, code, experimental materials, or other resources for reproducibility or re-use. This category includes explicit requests for supplementary information or links to data/code. & \textit{``the code or data availability weblink or vendor for the method''} \newline \textit{``I want TB expression system papers with downloadable protocols''} \newline \textit{``review papers listing open datasets for computational linguistics''} \\
\bottomrule
\end{tabular}
\caption{Complete ScholarQA search criteria taxonomy. Taxonomy was built through iterative human review and LLM-assisted analysis using GPT-4.1 on a sample of 1000 queries.}
\label{tab:sqa-criteria-full}
\end{table*}

\begin{table*}[t]
\centering
\footnotesize
\begin{tabular}{p{0.22\textwidth}p{0.38\textwidth}p{0.35\textwidth}}
\toprule
\textbf{Category} & \textbf{Description} & \textbf{Examples} \\
\midrule
Methodology-Specific Criteria & Queries where the user requires specific methods, approaches, analytical techniques, or experimental designs to be present in the search results. This can include demands for computational models, experimental paradigms, or meta-analyses. & \textit{``papers which use Nobel laureates' publication data to study interdisciplinary research''} \newline \textit{``Equilibrium analysis enables the assessment of whether observed outcomes reflect a stable strategic configuration...''} \newline \textit{``Find systematic reviews and/or clinical practice guidelines published in the last 10 years...''} \\
\addlinespace
Temporal Constraints & Queries that include requirements related to the publication date or period of the information (e.g., recent literature, post-2020, historical periods). This includes searching within specific timeframes or for recent advances. & \textit{``Find systematic reviews and/or clinical practice guidelines published in the last 10 years (2014–2025)...''} \newline \textit{``papers that have been published between 2023 and 2025''} \newline \textit{``papers about recent advances in communications sectors that Ofcom deals with''} \\
\addlinespace
Metadata-Based Criteria & Queries specifying attributes of the publications or their creators, such as language, authorship, institutional affiliation, location/country, or publication venue (journal, conference, etc.). & \textit{``Language: English or Portuguese.''} \newline \textit{``jurnal tentang Teori stimulus respons Neal E.Miller \& John Dollard''} \newline \textit{``In Sri Lanka, when a government employee is promoted...''} \\
\addlinespace
Publication Type/Quality Filters & Queries that require certain types of literature based on format, peer review status, or journal ranking, such as papers, theses, reviews, or guidelines, with or without requirements for high-impact or reputable sources. & \textit{``Find systematic reviews and/or clinical practice guidelines published in the last 10 years...''} \newline \textit{``any paper or thesis about nostalgia and self construal''} \newline \textit{``por favor me das una tesis de maestría de control constitucional''} \\
\addlinespace
Citation and Impact-Based Criteria & Queries that specify requirements related to citation counts, impact factors, highly-cited papers, or seminal works in a field. This can include searching for influential or foundational studies. & \textit{``papers which use Nobel laureates' publication data to study interdisciplinary research''} \newline \textit{``Find papers that refer to the relationships between degradation ability of PROTAC...''} \newline \textit{``Find papers that explore mechanisms that stimulate agent cooperation in MARL.''} \\
\addlinespace
Data/Resource Availability Criteria & Queries that ask for studies or papers with available datasets, code, experimental materials, or other resources for reproducibility or re-use. This category includes explicit requests for supplementary information or links to data/code. & \textit{``the code or data availability weblink or vendor for the method''} \newline \textit{``I want TB expression system papers with downloadable protocols''} \newline \textit{``review papers listing open datasets for computational linguistics''} \\
\bottomrule
\end{tabular}
\caption{Complete Paper Finder search criteria taxonomy. Taxonomy was built through iterative human review and LLM-assisted analysis using GPT-4.1 on a sample of 1000 queries.}
\label{tab:pf-criteria-full}
\end{table*}

\begin{table*}[t]
\centering
\footnotesize
\begin{tabular}{p{0.22\textwidth}p{0.38\textwidth}p{0.35\textwidth}}
\toprule
\textbf{Category} & \textbf{Description} & \textbf{Examples} \\
\midrule
Keyword-style Query & Short, often fragmentary, queries resembling search engine keywords or subject headings. No verbs or complete sentences; typically a list of nouns or concepts separated by spaces or simple punctuation. & \textit{``sludge production MBBR''} \newline \textit{``obsessive-compulsive personality''} \newline \textit{``hydrogen production methods''} \\
\addlinespace
Natural Language Question & Fully formed, grammatically complete questions using natural language. & \textit{``What is the role of agile human resource management in strategic entrepreneurship? The moderating role of corporate culture...''} \newline \textit{``How can more effective research on Neuromarketing be conducted?''} \newline \textit{``What are the most common ways (recently) to correct ocr errors''} \\
\addlinespace
Explicit Instruction/Imperative & Direct commands or requests instructing the system to perform an action (e.g., `find...', `review...', `compare...'). May be simple or include detailed steps, but typically begins with an imperative verb. & \textit{``Review papers on SOTA and research gaps in manifold learning in AI.''} \newline \textit{``Find policies and events in European countries that demonstrate `strategic autonomy' between 2016 and 2025.''} \newline \textit{``Compare las membranas poliméricas de inclusión respecto a cada uno de los métodos convencionales...''} \\
\addlinespace
Complex Contextual Narrative & Lengthy, detailed queries that provide substantial background context, motivations, definitions, or examples before asking a main question or issuing a command. These queries resemble a mini-narrative, sometimes including an abstract, data, citations, or technical context. & \textit{``In this experiment, various culture systems for the GF677 rootstock were compared... Is there an explanation in the scientific literature for why leaf number and leaf area are greater...''} \newline \textit{``I have four papers and I want to combine them all to be a doctoral dissertation...''} \newline \textit{``The performance of T-junction micromixers, as analysed in CFD models...''} \\
\addlinespace
Multi-part/Multi-step Query & Queries composed of multiple distinct sub-questions, tasks, or steps, often divided by letters, numbers, or separate sentences. The sub-queries are related but require separate or structured responses. & \textit{``Review papers on SOTA and research gaps in manifold learning in AI. For each of the top 9 methods extract: a) Key architectural or algorithmic ideas... b) Reported SOTA metric... c) the code or data availability...''} \newline \textit{``Describe what this cognitive ability encompasses... Please use academic papers... for each part of your response.''} \newline \textit{``Find me the research papers about interictal epileptiform discharges (IEDs)... Does STD in excitatory neurons...''} \\
\addlinespace
Boolean or Logical Operators & Queries using explicit logical or Boolean operators or patterns (AND, OR, NOT, +, -, parentheses, slashes, etc.) to combine multiple search terms, constrain results, or designate alternatives. Sometimes appears as `Topic A/Topic B', `x, y AND z'. & \textit{``adulteration detection in milk AND spectroscopy''} \newline \textit{``machine learning OR deep learning applications in medical imaging''} \newline \textit{``neural network NOT convolutional''} \\
\addlinespace
Citation/Format Specification & Queries where the format or style of the answer is explicitly specified, such as requiring specific citation styles (APA, etc.), in-text citations, references, or structured bibliographies. & \textit{``use APA style citation apply in-text citation and write what the meaning of `innovation'''} \newline \textit{``Please use academic papers and books when formulating your response. Ensure to provide references...''} \newline \textit{``Include appropriate references.''} \\
\bottomrule
\end{tabular}
\caption{Complete ScholarQA query phrasing taxonomy. Taxonomy was built through iterative human review and LLM-assisted analysis using GPT-4.1 on a sample of 1000 queries.}
\label{tab:sqa-phrasing-full}
\end{table*}

\begin{table*}[t]
\centering
\footnotesize
\begin{tabular}{p{0.22\textwidth}p{0.38\textwidth}p{0.35\textwidth}}
\toprule
\textbf{Category} & \textbf{Description} & \textbf{Examples} \\
\midrule
Keyword-style Query & Short, often fragmentary, queries resembling search engine keywords or subject headings. No verbs or complete sentences; typically a list of nouns or concepts separated by spaces or simple punctuation. & \textit{``algorithmic trading''} \newline \textit{``ppf''} \newline \textit{``bayesian optimization applications''} \\
\addlinespace
Natural Language Question & Fully formed, grammatically complete questions using natural language. & \textit{``How does fear of missing out influence the effect of mindfulness on academic procrastination among students?''} \newline \textit{``Do a literature review on artificial intelligence in management accounting''} \newline \textit{``What cognitive strategies and techniques can improve focus and concentration for artists...''} \\
\addlinespace
Explicit Instruction/Imperative & Direct commands or requests instructing the system to perform an action (e.g., `find...', `review...', `compare...'). May be simple or include detailed steps, but typically begins with an imperative verb. & \textit{``Find papers that refer to the relationships between degradation ability of PROTAC and potency or affinity...''} \newline \textit{``Give papers on finetuning an llm for writing research papers''} \newline \textit{``Find systematic reviews and/or clinical practice guidelines published in the last 10 years...''} \\
\addlinespace
Complex Contextual Narrative & Lengthy, detailed queries that provide substantial background context, motivations, definitions, or examples before asking a main question or issuing a command. These queries resemble a mini-narrative, sometimes including an abstract, data, citations, or technical context. & \textit{``As restored ecosystems evolve towards the composition and structure characteristic of native vegetation...''} \newline \textit{``Equilibrium analysis enables the assessment of whether observed outcomes reflect a stable strategic configuration...''} \newline \textit{``Papers about innovation and hurdles within industries that prevent it...''} \\
\addlinespace
Boolean or Logical Operators & Queries using explicit logical or Boolean operators or patterns (AND, OR, NOT, +, -, parentheses, slashes, etc.) to combine multiple search terms, constrain results, or designate alternatives. Sometimes appears as `Topic A/Topic B', `x, y AND z'. & \textit{``Find systematic reviews and/or clinical practice guidelines published in the last 10 years (2014–2025)...''} \newline \textit{``Papers introducing a dataset of an unscripted dialogue between 2 speakers...''} \newline \textit{``Stretching exercises are commonly integrated into physical education programs...''} \\
\addlinespace
Multi-part/Multi-step Query & Queries composed of multiple distinct sub-questions, tasks, or steps, often divided by letters, numbers, or separate sentences. The sub-queries are related but require separate or structured responses. & \textit{``Review papers on SOTA and research gaps in manifold learning in AI. For each of the top 9 methods extract: a) Key architectural or algorithmic ideas... b) Reported SOTA metric... c) the code or data availability...''} \newline \textit{``Describe what this cognitive ability encompasses... Please use academic papers... for each part of your response.''} \newline \textit{``Find me the research papers about interictal epileptiform discharges (IEDs)... Does STD in excitatory neurons...''} \\
\addlinespace
Citation/Format Specification & Queries where the format or style of the answer is explicitly specified, such as requiring specific citation styles (APA, etc.), in-text citations, references, or structured bibliographies. & \textit{``use APA style citation apply in-text citation and write what the meaning of `innovation'''} \newline \textit{``Please use academic papers and books when formulating your response. Ensure to provide references...''} \newline \textit{``Include appropriate references.''} \\
\bottomrule
\end{tabular}
\caption{Complete Paper Finder query phrasing taxonomy. Taxonomy was built through iterative human review and LLM-assisted analysis using GPT-4.1 on a sample of 1000 queries.}
\label{tab:pf-phrasing-full}
\end{table*}

\begin{table}[t]
\centering
\footnotesize
\begin{tabular}{ll}
\toprule
\textbf{Field of Study} & \textbf{Field of Study} \\
\midrule
Agricultural Sciences & Law and Legal Studies \\
Anthropology & Linguistics \\
Arts and Design & Literature \\
Biomedical Sciences & Mathematics \\
Biology & Mechanical Engineering \\
Business and Management & Philosophy \\
Chemistry & Physics \\
Civil Engineering & Political Science \\
Clinical Medicine & Psychology \\
Computer Science & Public Health \\
Earth Sciences & Sociology \\
Economics & Statistics \\
Education and Pedagogy & Environmental Studies \\
Electrical Engineering & History \\
\bottomrule
\end{tabular}
\caption{The 28 fields of study used for query classification.}
\label{tab:fields}
\label{app:fields}
\end{table}

%% file: appendix_prompts.tex
\section{LLM Prompts}
\label{app:prompts}

This section documents the prompts used for LLM-based classification and analysis tasks in this paper. Each prompt is accompanied by the Pydantic schema used for structured output decoding.

\subsection{Query Complexity Analysis}
\label{app:prompt-complexity}

This prompt extracts structural components (clauses, constraints, entities, and relations) from user queries to measure query complexity.


\begin{lstlisting}[basicstyle=\ttfamily\small,breaklines=true]
Analyze the following query and decompose it into its structural components:

Query: "{query}"

Please identify:
1. **Clauses**: Independent clauses or main statements in the query
   (e.g., separate questions, commands, or statements)
2. **Constraints**: Specific constraints or criteria mentioned
   (e.g., date ranges, publication venues, author names,
   methodological requirements)
3. **Entities**: Named entities or concepts referenced
   (e.g., specific theories, methods, diseases, technologies,
   people, places)
4. **Relations**: Relationships between entities that are mentioned
   or implied (e.g., "X influences Y", "A is a type of B",
   "C causes D")

For each category, provide a list of distinct items. If a category
doesn't apply to the query, return an empty list.
\end{lstlisting}

\textbf{Output Schema (Pydantic):}
\begin{lstlisting}[language=Python,basicstyle=\ttfamily\small]
class QueryComplexityAnalysis(BaseModel):
    clauses: List[str] = Field(
        description="List of independent clauses in the query"
    )
    constraints: List[str] = Field(
        description="List of constraints or criteria specified"
    )
    entities: List[str] = Field(
        description="List of named entities referenced in the query"
    )
    relations: List[str] = Field(
        description="List of relations between entities"
    )
\end{lstlisting}

\clearpage

\subsection{User Feedback Classification}
\label{app:prompt-feedback}

This prompt classifies user feedback into predefined categories to understand the types of issues and suggestions users report.


\begin{lstlisting}[basicstyle=\ttfamily\small,breaklines=true]
You are classifying user feedback for a research assistant tool
into predefined categories. A single feedback can belong to
MULTIPLE categories if applicable.

Available Categories:
{categories_description}

Feedback to classify:
"{feedback_text}"

Which categories apply to this feedback? Return a list of
category names. If no categories apply, return an empty list.
\end{lstlisting}

\textbf{Output Schema (Pydantic):}
\begin{lstlisting}[language=Python,basicstyle=\ttfamily\small]
class FeedbackClassification(BaseModel):
    category_names: List[str] = Field(
        description="List of applicable category names for the
                     feedback (multi-label)"
    )
\end{lstlisting}

\subsection{Duplicate Query Detection}
\label{app:prompt-duplicate}

This prompt identifies duplicate queries from a user's query history to understand query reuse patterns.


\begin{lstlisting}[basicstyle=\ttfamily\small,breaklines=true]
You are analyzing a user's search queries to identify duplicates.
Two queries should be considered duplicates if they are essentially
the same despite minor differences such as:
- Typos or spelling variations
- Word order changes
- Minor spacing or punctuation differences
- Slight phrasing or word choice changes that don't alter the
  core question

Here are the user's queries (numbered):

{numbered_queries}

Your task:
1. Group together queries that are duplicates
   (ones asking essentially asking the same thing)
2. Return the EXACT query text (not the numbers) in each group
3. Queries that are unique (no duplicates) should be in groups
   by themselves
4. Every query must appear exactly once in your output

Return your answer as JSON with a single key `groups` which is a
list of lists of duplicate queries.

Example:
If queries were:
1. What is machine learning?
2. what is machine learning
3. Define neural networks
4. What are neural networks?

You would return groups like:
{"groups":
    [
      ["What is machine learning?", "what is machine learning"],
      ["Define neural networks", "What are neural networks?"]
    ]
}
\end{lstlisting}

\textbf{Output Schema (Pydantic):}
\begin{lstlisting}[language=Python,basicstyle=\ttfamily\small]
class DuplicateQueryGroups(BaseModel):
    groups: List[List[str]] = Field(
        description="List of duplicate groups. Each group contains
                     the exact query texts that are duplicates of
                     each other. Single queries (no duplicates)
                     should be in groups of size 1. ALL input
                     queries must appear exactly once."
    )
\end{lstlisting}

\clearpage

\subsection{Response Quality Assessment}
\label{app:prompt-quality}

This prompt assesses the quality of AI-generated responses using an LLM-as-judge approach.


\begin{lstlisting}[basicstyle=\ttfamily\small,breaklines=true]
You are evaluating the quality of an AI-generated response to a
user's query.

User Query: {query}

AI Response:
{formatted_report}

Based on the relevance, accuracy, completeness, and usefulness of
this response, would you give it a thumbs up (good quality) or
thumbs down (poor quality)?

Consider:
- Does the response adequately address the user's query?
- Is the information relevant and well-organized?
- Would a typical user find this response helpful?
\end{lstlisting}

\textbf{Output Schema (Pydantic):}
\begin{lstlisting}[language=Python,basicstyle=\ttfamily\small]
class FeedbackAssessment(BaseModel):
    assessment: str = Field(
        description="Assessment of the response quality:
                     'thumbs_up' or 'thumbs_down'"
    )
    reasoning: str = Field(
        description="Brief reasoning for the assessment
                     (1-2 sentences)"
    )
\end{lstlisting}

\clearpage

\subsection{Query Intent Classification}
\label{app:prompt-intent}

This prompt classifies user queries into intent categories to understand the underlying purpose of each query.


\begin{lstlisting}[basicstyle=\ttfamily\small,breaklines=true]
Classify the following academic query into one or more intent
categories from the provided list. A query can belong to multiple
categories if it has multiple purposes.

Query: "{query}"

Available intent categories:
{intent_list}

Please respond with a JSON object containing a list of applicable
intent names. Only include intent names that clearly apply to
this query.
\end{lstlisting}

\textbf{Output Schema (Pydantic):}
\begin{lstlisting}[language=Python,basicstyle=\ttfamily\small]
class MultilabelQueryIntentResult(BaseModel):
    intent_names: List[str] = Field(
        description="List of applicable intent names for the query"
    )
\end{lstlisting}

\subsection{Query Phrasing Classification}
\label{app:prompt-phrasing}

This prompt classifies queries based on their phrasing patterns and style (e.g., keyword-style, natural language question, imperative command).


\begin{lstlisting}[basicstyle=\ttfamily\small,breaklines=true]
Classify the following academic query based on its phrasing
patterns and style. A query can exhibit multiple phrasing styles -
select ALL categories that apply.

Query: "{query}"

Available phrasing categories:
{category_list}

Please respond with a JSON object containing a list of ALL
applicable phrasing category names. Only include categories that
clearly match the phrasing style of this query.
\end{lstlisting}

\textbf{Output Schema (Pydantic):}
\begin{lstlisting}[language=Python,basicstyle=\ttfamily\small]
class QueryPhrasingResult(BaseModel):
    phrasing_types: List[str] = Field(
        description="List of applicable phrasing category names
                     for the query"
    )
\end{lstlisting}

\subsection{Field of Study Classification}
\label{app:prompt-field}

This prompt classifies queries into academic fields of study.


\begin{lstlisting}[basicstyle=\ttfamily\small,breaklines=true]
Classify the following academic query into one or more fields of
study from the provided list. A query can belong to multiple
fields if it spans multiple disciplines.

Query: "{query}"

Available fields of study:
{fields_list}

Please respond with a JSON object containing a list of applicable
field names. Only include field names that clearly apply to this
query. Focus on the primary discipline involved and only include
more than one field if the query is truly interdisciplinary and
not covered by a single field. Use the exact field names from the
provided list.
\end{lstlisting}

\textbf{Output Schema (Pydantic):}
\begin{lstlisting}[language=Python,basicstyle=\ttfamily\small]
class MultilabelFieldClassificationResult(BaseModel):
    field_names: List[str] = Field(
        description="List of applicable field of study names
                     for the query"
    )
\end{lstlisting}

\subsection{Search Criteria Classification}
\label{app:prompt-criteria}

This prompt classifies queries based on the types of search criteria specified (e.g., temporal constraints, methodology requirements, publication type filters).


\begin{lstlisting}[basicstyle=\ttfamily\small,breaklines=true]
Classify the following academic query based on the types of search
criteria being used. A query can combine multiple search criteria
types - select ALL categories that apply.

Query: "{query}"

Available search criteria categories:
{category_list}

Please respond with a JSON object containing a list of ALL
applicable criteria category names. Only include categories that
clearly match the search criteria used in this query.
\end{lstlisting}

\textbf{Output Schema (Pydantic):}
\begin{lstlisting}[language=Python,basicstyle=\ttfamily\small]
class SearchCriteriaResult(BaseModel):
    criteria_types: List[str] = Field(
        description="List of applicable criteria category names
                     for the query"
    )
\end{lstlisting}